\documentclass{article}

\usepackage{arxiv}

\usepackage[utf8]{inputenc} % allow utf-8 input
\usepackage[T1]{fontenc}    % use 8-bit T1 fonts
\usepackage{hyperref}       % hyperlinks
\usepackage{url}            % simple URL typesetting
\usepackage{booktabs}       % professional-quality tables
\usepackage{amsfonts}       % blackboard math symbols
\usepackage{nicefrac}       % compact symbols for 1/2, etc.
\usepackage{microtype}      % microtypography
\usepackage{cleveref}       % smart cross-referencing
\usepackage{lipsum}         % Can be removed after putting your text content
\usepackage{graphicx}
\usepackage{natbib}
\usepackage{doi}
\usepackage{listings}
\usepackage{color}
\usepackage{xcolor}

\newcommand{\CheckmarkBold}{\includegraphics{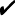}}
\newcommand{\highimpact}{\includegraphics{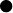}}
\newcommand{\medimpact}{\includegraphics{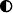}}
\newcommand{\lowimpact}{\includegraphics{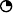}}

\definecolor{gray}{rgb}{0.4,0.4,0.4}
\definecolor{darkblue}{rgb}{0.0,0.0,0.6}
\definecolor{cyan}{rgb}{0.0,0.6,0.6}
\definecolor{maroon}{rgb}{0.5,0,0}
\definecolor{darkgreen}{rgb}{0,0.5,0}
\definecolor{OrangeRed}{HTML}{613F99}

\hypersetup{
    colorlinks=true,
    allcolors=blue,
    pdftitle={Digital requirements engineering with an INCOSE-derived SysML meta-model},
    pdfauthor = {James S. Wheaton and Daniel R. Herber},
    }

\newsavebox{\mybox}
\newcommand{\Stereotype}[1]{\guillemotleft{}{#1}\guillemotright{}}

\newcommand{\SA}[1]{#1}
\newcommand{\SB}[1]{#1}

\title{Digital requirements engineering with an INCOSE-derived SysML meta-model}

\author{ \href{https://orcid.org/0009-0000-3553-908X}{\includegraphics[scale=0.06]{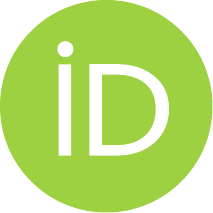}\hspace{1mm}James S.~Wheaton} \\
	Department of Systems Engineering\\
	Colorado State University\\
	Fort Collins, Colorado, USA \\
	\texttt{jsw@cybersystems.engineer} \\
	\And
	\href{https://orcid.org/0000-0003-4995-7375}{\includegraphics[scale=0.06]{orcid.pdf}\hspace{1mm}Daniel R.~Herber} \\
	Department of Systems Engineering\\
	Colorado State University\\
	Fort Collins, Colorado, USA \\
	\texttt{daniel.herber@colostate.edu} \\
}

\renewcommand{\shorttitle}{J.~S.~Wheaton \& D.~R.~Herber}

\begin{document}
\maketitle

\begin{abstract}%
%<*mytagR16b>
Traditional requirements engineering tools do not readily access the
SysML-defined system architecture model, often resulting in ad-hoc
duplication of model elements that lacks the connectivity and expressive
detail possible in a \SB{Systems Modeling Language (SysML)}-defined model.
%</mytagR16b> 
Further integration of requirements engineering activities
with MBSE contributes to the Authoritative Source of Truth while
facilitating deep access to system architecture model elements for V\&V activities.
%<*mytagR5C6>
\SA{Building from prior work on} Model-Based Structured Requirement\SA{s, we extend \SB{its}}
SysML profile to comply with the INCOSE \emph{Guide to Writing Requirements}
while conforming to the ISO/IEC/IEEE 29148 standard requirement statement patterns.
%</mytagR5C6>
Rules, Characteristics, and Attributes were defined in SysML according to the \emph{Guide} to facilitate
requirements \SA{development}, verification \& validation.
The resulting SysML \SA{p}rofile was applied in two system architecture models at NASA Jet
Propulsion Laboratory, allowing us to assess its applicability\SA{,}
value\SA{, and gaps} in real-world project environments. Initial results indicate that \SA{use of the}
INCOSE-derived Model-Based Structured Requirement \SA{profile} may rapidly improve
requirement expression quality while complementing the \emph{NASA
Systems Engineering Handbook} checklist and guidance, but typical
requirement management activities still have challenges related to
automation and support in the \SA{SysML} modeling software.
\end{abstract}

% Please include a maximum of seven keywords

%%%%%%%%%%%%%%%%%%%%%%%%%%%%%%%%%%%%%%%%%%%%%%%%%%%%%%

\clearpage

\section{Introduction}
\label{sec:Intro}

As engineered systems grow in complexity, the demand for more cost-effective system development programs grows in turn. A critical point of leverage in reducing system development costs is by improving requirements engineering processes and the quality of its outputs \citep{NASA_SE_HB2017}. Digital engineering (DE) promises to improve quality and reduce costs through increased access and connectivity of digital artifacts in a central data store called the \SB{authoritative source of truth} (ASoT) \citep{noguchi2020digital}. \SB{Model-based systems engineering} (MBSE) improves upon \SA{document-centric systems engineering} by incorporating formal digital models throughout systems engineering processes and products, by leveraging those digital system models to precisely represent design values and relationships, and by computing model validity based on modeling language and design rules. MBSE practice develops digital requirements engineering (DRE) as a response to increasing the system architecture model connectivity toward complete traceability of stakeholder needs and system requirements in support of DE goals.

%<*mytagR3C2>
\SA{With MBSE being central to the INCOSE systems engineering vision dating back to 2007 \citep{INCOSEvision2020} and DE later being added in 2014 \citep{INCOSEvision2025}, and carried forward to the 2035 Vision \citep{INCOSEvision2035}, systems engineers should apply effort to incorporate model-based, digital methods in their practice.}
The effectiveness of MBSE compared to \SA{document-centric systems engineering continues to be studied, but results already corroborate its expected benefits} \citep{rogers2021mbse,younse2021mbse,systems7010012,osti_1561164},
making it the most important systems engineering practice toward achieving DE goals in the organization.
%</mytagR3C2>
As a motivating example, the NASA-ESA Mars Sample Return \SA{campaign} is ``an ambitious and complex space
system engineering endeavor'' \citep{sundararajan2022understanding} with multiple
interfacing space systems necessary to coordinate the safe return of Martian gas and solid core samples for further study on Earth.
The Mars Sample Return \SA{program} \SA{completed} its second Independent Review Board
assessment that include\SA{d} a probable program life cycle cost estimate of
US\$8-11 billion, ``strong irrefutable evidence'' that strong systems
engineering (SE) is a crucial factor for mission success, and recommendations
to refactor the program architecture to control costs \citep{MSR_IRB2023}.
%<*mytagR3C3>
Strong \SA{systems engineering} is increasingly model-based, and \SA{while DRE is not a substitute for good requirements engineering practice,
it highlights a paradigm shift that includes other digital engineering trends.}
%</mytagR3C3>

\subsection{Digital Engineering}
\label{sec:DE}

%<*mytagR5C7>
Recognition of the current and potential impact of digital models,
including those used in MBSE, has \SB{led the US Department of Defense to develop} a strategy
\SB{that makes} digital models \SB{essential to} the system development process.
%</mytagR5C7>
DE is ``an integrated digital approach that uses
authoritative sources of system data and models as a continuum across
disciplines to support lifecycle activities from concept through
disposal'' \citep{DoD_DE_Strategy2018}. As an engineering leader,
NASA has invited the future of digital workflows by publishing a
Digital Transformation strategy \citep{marlowe2022}, MBSE strategy \citep{weiland2021},
and DE Acquisition Framework Handbook \citep{NASA_DE_HDBK}.
DE is not a new discipline of engineering but rather an intentional transformation of how an organization
integrates and performs its engineering activities to achieve higher
quality and efficiency \citep{noguchi2020digital}.

One of the DE goals is to provide an enduring ASoT
of the system to improve communication and decision-making. The system architecture model is one
component of the ASoT, typically integrated in a centralized repository,
and the system requirements may be created in the architecture model or
synchronized with the model from a \SB{requirements management tool} (RMT).
DRE further integrates requirements with
the ASoT, enabling formal verification and validation (V\&V) activities
that may be automated to improve model confidence and ease stakeholder
reviews \citep{duprez2023approach}.

\subsection{Requirements Engineering}
\label{sec:RE}

%<*mytagR5C9>
Requirements engineering (RE) is a \SB{process area} of systems engineering that
encompasses requirements development and requirements management.
%</mytagR5C9>
The ISO/IEC/IEEE 29148:2018 standard defines requirements engineering as
``an interdisciplinary function that mediates between the domains of the
acquirer and supplier or developer to establish and maintain the
requirements to be met by the system, software or service of interest.
Requirements engineering is concerned with discovering, eliciting,
developing, analyzing, verifying (including verification methods and
strategy), validating, communicating, documenting and managing
requirements'' \citep{ISO29148}. The range of RE activities necessitates the use of metadata
to organize information about each requirement, emphasizing that the familiar ``shall'' statement is only
one attribute of a well-managed and model-connected requirement.

The INCOSE \emph{Guide to Writing Requirements} (GtWR) provides a
current perspective of well-formed requirements, and it defines a
requirement statement as ``the result of a formal transformation of one
or more sources, needs, or higher-level requirements into an agreed-to
obligation for an entity to perform some function or possess some
quality within specified constraints with acceptable risk'' \citep{GtWR2023}.
The GtWR emphasizes that the requirement statement
forms the basis of contractual language, and then presents a rules-based
structured format for facilitating that communication. It defines a
requirement expression as the requirement statement and its attributes. The GtWR recommends a
data-centric practice using a RMT, as opposed to spreadsheets or documents, to model and present
requirement expressions using diagrams and tables for stakeholder-tailored views.

Systems engineering handbooks provide another important reference for
requirements engineering activities, complementing the detailed guides,
manuals, and standards cited above. The INCOSE \textit{Systems Engineering
Handbook - Fifth Edition} provides updated Sections 2.3.5.2 and 2.3.5.3
that incorporate the latest INCOSE guides and manuals on needs and
requirements engineering \citep{INCOSE_SE_HB2023}. The NASA Systems Engineering
Handbook \citep{NASA_SE_HB2017} describes the traditional
NASA requirements definition and management processes in Sections 4.2
and 6.2, respectively, emphasizing bidirectional traceability, and
including a checklist in Appendix C and an informal set of
characteristics similar to those defined in the latest INCOSE GtWR. The
INCOSE GtWR and NASA sets of characteristics are compared in Sec. \ref{sec:Discussion} and
found to be complementary.

\subsection{Systems Modeling Language}
\label{sec:SysML}

The Systems Modeling Language (SysML) by the Object Management Group (OMG) \citep{SysML1.7} is a
standard language for modeling system architectures which provides the
capability to model the solution space as structure, behavior, rules 
and requirements as digital model elements in a directional graph 
constrained by SysML semantics. SysML version 1.7 is expected to be the final version in the 1.x series
as the standards development effort shifts to the new version 2.
SysML provides rudimentary facilities for modeling system requirements,
including the primary attributes: ID, name, and text; and the
relationships: derive, refine, satisfy, verify, and trace (which is
discouraged in favor of the more precise relationships). Requirement
type and rationale are not attributes provided by the SysML standard but
are customizations of the SysML profile often provided by systems architecture
modeling tools. According to the standard, SysML Requirements may be
shown in a Requirements Diagram, or placed on other SysML diagrams to
highlight relationships for certain stakeholder views; requirements
tables and matrices are non-normative.

Today RE is often practiced with the help of an RMT
which stores the requirements in a database and provides structured 
access to them for management activities. Since this RE practice uses 
digital models in a computer database, it may be considered MBSE or DE.
However, the threat remains of duplicative system model elements interfering with traceability due to the RMT lacking the SysML meta-model used to define the system architecture. Controlled import/export or RMT data connector synchronization cycles alleviate this problem but may
cause issues due to the non-standard interfaces between tools \citep{Wheaton2024a}.
%<*mytagR3C5>
Although SysML v2 has addressed this issue by standardizing the API 
providing access to the system architecture model \SA{\citep{bajaj2022sysmlv2}}, \SA{due to disparate tool development lifecycles and enterprise adoption timelines,} it will be years before it sees widespread adoption inside \SA{organizations \citep{Call2022b}}.
%</mytagR3C5>
SysML v2 improves upon v1 by modeling requirements as constraints that must be satisfied by \SA{parts of} the system-of-interest,
and it uses the consistent distinction between a requirement definition and requirement
usage to aid in reuse \citep{SysMLv2}.

%<*mytagR3C6>
%<*mytagR1C2-0>
Due to the perceived inadequate facilities for modeling and managing
requirements \SA{using SysML}, \SA{the GtWR cautions against its use} %SysML has not been favored by requirements engineers
\citep{GtWR2023}.
%</mytagR1C2-0>
Its apparent advantages with respect to its
graphical syntax (diagrams) \SA{and traceability support} have not been enough to satisfy the critical
needs of managing and distributing possibly thousands of requirements.
\SA{In addition, due to accessibility and training challenges associated with SysML modeling tools,
the GtWR notes that stakeholders ``will still prefer and demand'' electronic/printed documents ``for the foreseeable future''. }
However, SysML makes it possible to extend the language through Stereotypes, thus
availing the systems engineer with meta-modeling capabilities to define
custom elements and to use them \SA{to model RE concepts directly.} %as just another SysML element.
\SA{SysML tools also provide means to generate artifacts such as requirements documents from the architecture model,
ameliorating the need for specialized software for stakeholder review.}
%<*mytagR4C4-1>
This \SA{SysML} meta-modeling capability \SA{specifically} enables the \SA{GtWR-derived} Model-Based Structured Requirement (MBSR)
approach described in this paper\SA{, while model-based artifact generation enabled benefits and encountered challenges discussed in Section \ref{sec:Discussion}}.
%</mytagR4C4-1>
%</mytagR3C6>

\subsection{Overview}
\label{sec:Overview}

%<*mytagOverview>
This paper presents an extension to the MBSR approach developed by \citet{herber2023} and \citet{herber2022model}
that incorporates the INCOSE GtWR ontology and the \citet{ISO29148} standard pattern for requirement statements
(Fig. \ref{fig:Metamodel})\SB{,} using a notional space flight system to demonstrate the use of the SysML profile.
%<*mytagR4C4-2>
\SB{Model-based application of the GtWR represents a gap in the literature that this paper aims to address.}
%</mytagR4C4-2>

%<*mytagR0C2>
%<*mytagR5C15>
\SA{\SB{The c}ontributions of this paper \SB{may be summarized as follows}:}
%</mytagR5C15>

%<*mytagR5C16>
\begin{itemize}
	\item \SA{We present a complete open-source SysML profile for digital requirements engineering based on structured requirement statements \SB{from ISO/IEC/IEEE 29148:2018 and formalization of the meta-model from INCOSE GtWR, not available from INCOSE nor similarly published by other research we reviewed}.}
    %<*mytagR5C17>
	\item \SA{We demonstrate \SB{the} application of the \SB{INCOSE-derived MBSR} profile using an example model in the aerospace domain, \SB{adding to the domains studied using MBSR, and describing additional supported capabilities with respect to DRE practice}.}
    %</mytagR5C17>
	\item \SA{We validate \SB{INCOSE-derived MBSR} in a real-world NASA/ESA Mars Sample Return project in Pre-Phase A \citep{NASA_SE_HB2017} and discuss the \SB{perceived} benefits and challenges.} \SB{This work shows how MBSR was applied early in the concept system lifecycle phase compared to late in the development phase which was covered in prior MBSR work. Readers in the aerospace domain may examine this evidence when considering to apply INCOSE-derived MBSR in their work.}
\end{itemize}
%</mytagR5C16>
%</mytagR0C2>

%<*mytagR3C11>
The rest of the paper is organized as follows: 
Sec.~\ref{sec:MBSR} \SA{describes the prior MBSR SysML language extension that connects model elements to their respective statement pattern slots;}
Sec.~\ref{sec:INCOSE-MBSR} \SA{extends the prior MBSR with our INCOSE-derived SysML meta-model and presents a representative example and summarized case study;}
Sec.~\ref{sec:Discussion} discusses the benefits and challenges of this approach \SA{compared to unstructured document-based and classical SysML requirements in the context of a real project in Pre-Phase A;}
\SA{Sec.~\ref{sec:RelatedWork} presents a literature review and compares SysML- and other model-based requirement techniques to INCOSE-derived MBSR;}
and Sec.~\ref{sec:Conclusion} concludes \SA{with limitations and future work, and} reflects on the \SA{value} in the context of DRE.
%</mytagR3C11>
%</mytagOverview>

%%%%%%%%%%%%%%%%%%%%%%%%%%%%%%%%%%%%%%%%%%%%%%%%%%%

\section{Model-Based Structured Requirement (MBSR)}
\label{sec:MBSR}
A strong SE practice is a countervailing force against cost overruns and system development project cancellations,
and it is bolstered by a strong embedded RE practice from the beginning of the project. Where there were undefined,
unsupported, or undisciplined RE processes in an organization, tailored guidance from INCOSE should fill the knowledge gap.
Developing successful complex systems requires participation from diverse sets of stakeholders and organizations, leading
to the development of Simplified Technical English \citep{ASDSTE100} to reduce language ambiguity and confusion.
Unlike the situation in the late 20th century when software development projects turned into production messes and
accumulated technical debt, practitioners today may now benefit from advancements in RE to minimize ambiguity, reduce technical debt,
formalize V\&V processes, and produce consistently high-quality design output specifications \citep{NRM2022,avdeenko2016ontology}.

%<*mytagR5C18>
This section of the paper describes \SB{classical SysML} requirements modeling and performs a gap analysis (Sec. \ref{sec:ClassicalSysML}), structured requirements and its related work (Sec. \ref{sec:StructuredReq}),
and INCOSE GtWR-derived meta-model extensions to MBSR from \citet{herber2023} (Sec. \ref{sec:MBSR}).
%</mytagR5C18>

\subsection{Classical SysML Requirements Modeling}
\label{sec:ClassicalSysML}

Classical SysML represents a requirement as an ``indivisible entity'' \citep{dick2017integrating} that enriches
and is enriched by other system architecture model elements through the use of typed relationships.
According to the SysML standard specification, ``a requirement is defined as a stereotype of UML Class subject to a set of constraints'' and includes ``properties to specify its unique identifier and text requirement,'' noting that ``additional properties ... can be specified by the user'' \citep{SysML1.7}. The `name', `id', and `text' properties of SysML requirements are defined as a simple string of characters, lacking any other structure, manipulable only by standard string processing functions of popular programming languages if available. Traceability relationships include containment, and subtypes of the UML Dependency relationship, as follows:
\begin{description}
\item[\textit{Containment}] specifies the \textit{Owner} of the requirement in the model containment hierarchy, graphically represented by a circle with two perpendicular lines crossed at the \textit{Owner} end of the connector.
\item[\textit{Derive}] specifies a type of \textit{Trace} that relates a derived requirement to its source requirement at the arrowhead-end of the dashed line.
\item[\textit{Refine}] specifies a directed relationship used ``to describe how a model element or set of elements can be used to further refine a requirement'' \citep{SysML1.7}.
\item[\textit{Satisfy}] specifies a type of \textit{Trace} that ``describes how a design or implementation model satisfies one or more requirements'' \citep{SysML1.7}.
\item[\textit{Verify}] specifies a type of \textit{Trace} that relates a test case or other model element to a requirement to identify a verification activity that checks that the system element meets its traced requirements and constraints.
\item[\textit{Copy}] specifies a type of \textit{Trace} that creates a read-only copy of the requirement ID and textual statement on the requirement element at the non-arrowhead end of the dashed line.
\item[\textit{Trace}] is a general-purpose relationship between a requirement and any other model element; its use with more specific traceability relationships listed above is discouraged due its ambiguity \citep{SysML1.7}.
\end{description}

%%%%
\begin{figure}[bt]
\centering
\includegraphics[width=\columnwidth]{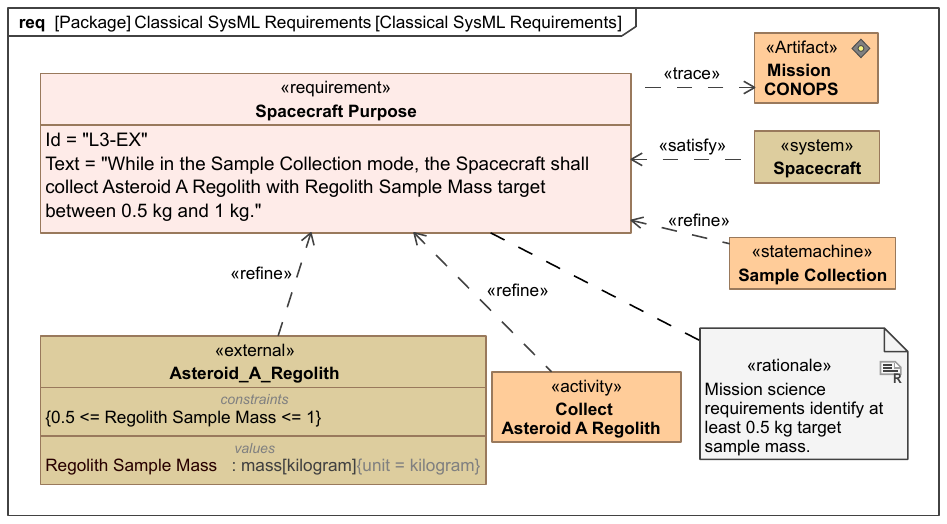}
\caption{Classical SysML requirements modeling using standard relationships to system model elements.}
\label{fig:ClassicalSysML}
\end{figure}
%%%%

Classical SysML-based requirements popularized the idea of ``cohabitation''
where requirements and system model elements exist and are linked together in
the same model \citep{bernard2012requirements} (Fig. \ref{fig:ClassicalSysML}).
Practicing DRE with this MBSE approach leads to the notion that requirements
engineers and system architects should work in integrated teams rather than
organizational silos and separate software tools \citep{NRM2022}.
%<*mytagR3C7>
However, due to the lack of requirements management facilities such as software-defined
workflows, audit logs, and change management boilerplate, managing requirements
in SysML \SA{presents additional challenges already addressed by enterprise-grade RMTs.}
\SA{While the GtWR notes that ``SysML requirement diagrams \dots~are not well-suited to representing multiple or large sets of requirements'' \citep{GtWR2023}, we find that the same may also apply to requirement tables compared to RMTs (see Sec. \ref{sec:Challenges}).}
%</mytagR3C7>

In order for system architects to access the system
requirements and add traceability relationships to the model, data connectors
with RMTs or import/export/sync with Excel spreadsheets are used.
Synchronization between tools mitigates the change management issues but due to
differing data schema and meta-models, important information enriching the
requirement expressions may not be available. Synchronized requirement tables
will automatically highlight requirements that have changed, but the implication
is that organizational silos are loosely cooperating in this manner, with the
RMT owning the requirement ASoT and the SysML tool working with copies of that data.

DRE has outgrown classical SysML requirements due to their lack of precision, leading
practitioners and tool vendors to extend standard SysML in various ways to satisfy MBSE objectives \citep{SysML1.7}.
The \textit{Trace} and \textit{Copy} relationships are discouraged, leading to confusion
about their use, although this may be mitigated by hiding them in the SysML tool's user perspective.
The \textit{Verify} relationship perpetuates the conflation of system V\&V with requirements V\&V (``verifying a requirement''),
a distinction made clear in the INCOSE GtWR. The non-normative extensions, such as the additional requirement stereotypes and risk kinds
in \citet{SysML1.7}, may confuse practitioners due to the `one size fits all' approach, and interface requirements, in particular, are rejected by the GtWR. Although SysML provides the basic utilities for complex
system requirements analysis, the implementation of such analyses can take considerable effort, favoring
organizations with mature model libraries. In summary, classical SysML requirements lack modeling precision
without significant extensions, and introduce ambiguity and confusion in modeling constructs
that contradict the latest RE guidance from INCOSE and ISO/IEC/IEEE.

\subsection{Structured Requirement}
\label{sec:StructuredReq}

The INCOSE GtWR and Needs and Requirements Manual (NRM) advocates for ``structured, natural language'' that treats the familiar `shall' statement not as an ``atomic entity'' but as a ``grammatical structure appropriate for communicating needs and requirements'' \citep{NRM2022}.
%<*mytagR3C10-1>
In fact, rule number one (R1) in GtWR is ``Structured Statement'' which contributes to the quality characteristics: (C3) unambiguous, (C4) complete, (C5) singular, (C7) verifiable, and (C9) conforming (refer to Table \ref{tab:Characteristics} and the GtWR Summary Sheet).
%</mytagR3C10-1>
%<*mytagR3C8>
This INCOSE guidance builds upon a history of success using structured language for textual requirements \citep{gilb2004,mavin2010big,dick2017integrating,carson2015implementing,carson2021}.

Requirement statements with uniform structure are shown to improve quality \SA{\citep{carson2015implementing,carson2021}} and have been adopted and recommended by \citet{ISO29148} and \citet{INCOSE_SE_HB2023}.
%</mytagR3C8>
Textual ``shall'' statements with parts in a standard order are easier to write, parse,
and verify due to the regular structure that guards against ambiguity and complex grammar. A structured requirement defines a pattern of the requirement statement using placeholders to clearly identify the critical features of the requirement. They have been called template requirements or ``statement-level templates'' \citep{GtWR2023}, structured requirements \citep{carson2015implementing}, requirement structures \citep{mavin2010big}, or ``boilerplates'' \citep{hull2010requirements}, but the GtWR uses the term `requirement pattern' to avoid confusion with other common uses of the word `template'. A requirement pattern is ``represented by a series of building blocks (also called pattern slots) including all the elements envisioned to represent a well-formed, singular, and complete requirement'' \citep{GtWR2023}. We present the following review of structured requirements using recommended patterns from \citet{carson2015implementing} and \citet{ISO29148}.
\begin{center}
\textit{Plain-language Requirement Pattern:}

The \textbf{{[}Who{]}} \textit{shall} \textbf{{[}What{]}} \textbf{{[}How Well{]}} under \textbf{{[}Condition{]}}.
\end{center}
\begin{description}
    \item[\textbf{{[}Who{]}}] Singular subject of the requirement referring to an entity or agent that provides a capability or performs a function.
    \item[\textbf{{[}What{]}}] Singular action-verb performed by the \textbf{{[}Who{]}}, referring to required functionality or quality characteristic.
    \item[\textbf{{[}How Well{]}}] Comparison factor(s) specified by constraints on the \textbf{{[}What{]}} which places feasible limits on the required functionality or quality characteristic, and which are used to verify the \textbf{{[}What{]}}.
    \item[\textbf{{[}Condition{]}}] ``[M]easurable qualitative or quantitative terms specified by characteristics such as an operational scenario, environmental condition, or a cause that is stipulated for a requirement'' \citep{herber2022model}.
\end{description}

\begin{center}
\textit{Plain-language Structured Requirement Example:}

The \textbf{{[}Spacecraft{]}} shall \textbf{{[}collect Asteroid\_A\_Regolith{]}} \textbf{{[}with Regolith\_Sample\_Mass target between 0.5 kg and 1 kg{]}} under \textbf{{[}Sample\_Collection mode{]}}.
\end{center}

The plain-language requirement pattern is flexible for use with functional requirements and quality requirements, and the pattern slot names may suit stakeholders who are not as comfortable with SE jargon.
%<*mytagR5C23>
\SB{Using} this requirement pattern consistently across requirement sets will significantly \SB{improve readability} compared to unstructured requirements.
%</mytagR5C23>
The \citet{ISO29148} standard pattern provides different names for the pattern slots and adds one for \textbf{{[}Object{]}} but is otherwise similar to the plain-language requirement pattern. The standard presents two patterns with an implied ``shall'' after the \textbf{{[}Subject{]}}:

\begin{center}
\textit{ISO/IEC/IEEE 29148:2018 Requirement Pattern:}

\textbf{{[}Subject{]} \textit{shall} {[}Action{]} {[}Constraint of Action{]}}.

\emph{OR}

\textbf{{[}Condition{]}, {[}Subject{]} \textit{shall} {[}Action{]} {[}Object{]}
{[}Constraint of Action{]}}.
\end{center}

\begin{description}
    \item[\textbf{{[}Condition{]}}] ``[M]easurable qualitative or quantitative attributes that are stipulated for a requirement, \dots and provide attributes that permit a requirement to be formulated and stated in a manner that can be validated and verified'' \citep{ISO29148}.
    \item[\textbf{{[}Subject{]}}] Singular system element in the same system hierarchy level as the requirement that provides a capability or performs a function.
    \item[\textbf{{[}Action{]}}] Singular action-verb performed by the \textbf{{[}Subject{]}}, referring to required functionality or quality characteristic.
    \item[\textbf{{[}Object{]}}] The entity being acted upon by the \textbf{{[}Subject{]}}; an element of the system or system environment.
    \item[\textbf{{[}Constraint of Action{]}}] The ``measurable outcome'' \citep{GtWR2023} that ``restrict[s] the design solution or implementation of the systems engineering process'' \citep{ISO29148} by applying feasible limits on the \textbf{[Action]} performed by the \textbf{[Subject]} on the \textbf{[Object]} under the stated \textbf{[Condition]}.
\end{description}

\begin{center}
\textit{ISO-standard Structured Requirement Example:}\\

\textbf{{[}While in the Sample\_Collection mode{]}}, the \textbf{{[}Spacecraft{]}} shall \textbf{{[}collect{]}} \textbf{{[}Asteroid\_A\_Regolith{]}} \textbf{{[}with Regolith\_Sample\_Mass target between 0.5 kg and 1 kg{]}}.
\end{center}

Many patterns are possible depending on the domain and should be prescribed by the practicing organization
(see Appendix C of GtWR for more examples). Here the Subject refers to the part of the system corresponding to the same
level as the requirement. The `shall' keyword has been inserted above to clarify the pattern
and to emphasize that ``requirements are mandatory binding provisions and use `shall''' \citep{ISO29148}.
The Action signifies that the Subject \emph{does} something --- `shall not' is forbidden (R16 in GtWR) --- and
that the statement is written in the active voice (R2 in GtWR),
avoiding superfluous and possibly confusing verbiage such as ``be capable of'' (R10 in GtWR).
This pattern makes clear that every requirement must have a verifiable Constraint.
Although every requirement has an associated condition of when it is
active, the first pattern may be used in the high-level functional
requirements when the Condition is ``ubiquitous'' \citep{GtWR2023}.
The latter pattern may be used as a default value for a
redefined `Text' SysML Property.
MBSE practice encourages the use of these pattern slots to reference system model elements,
not just defined terms in a project glossary, while RE practice as discussed in the GtWR emphasizes the important
role of textual requirement statements especially in the presence of system model views used to enhance stakeholder understanding.

\subsection{Plain-Language MBSR}
\label{sec:Plain-languageMBSR}

\begin{figure}[p]
\centering
\includegraphics[width=\columnwidth]{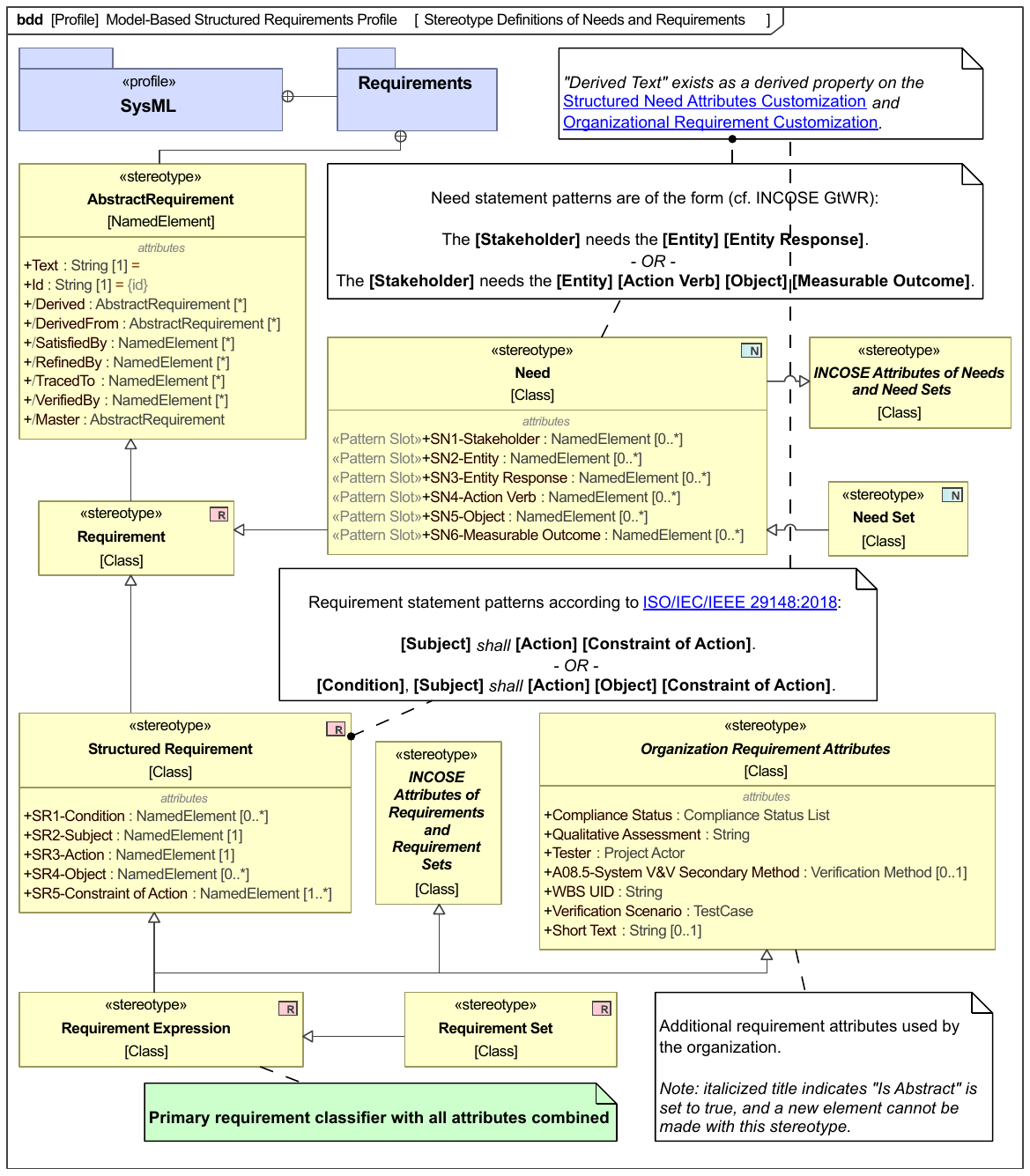}
\caption{SysML Stereotype definitions for Requirement Expression and Requirement Set. Refer to Fig. \ref{fig:GtWR-Attributes} for further detail on the INCOSE GtWR Attributes.}
\label{fig:Stereotypes}
\end{figure}

\begin{figure}[bt]
\centering
\includegraphics[width=\columnwidth]{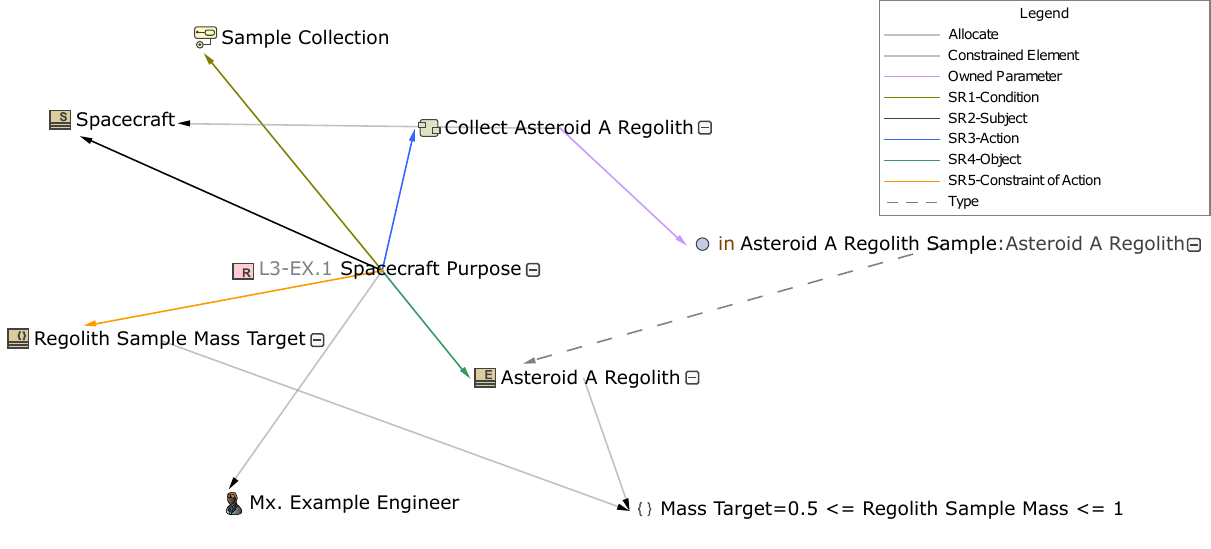}
\caption{Relation Map of example MBSR and related system architecture SysML elements.}
\label{fig:RelationMap}
\end{figure}

MBSR is an adaptation of structured requirements to the MBSE paradigm developed by \citet{herber2023} and \citet{herber2022model}, building on the pattern concept by making the pattern slots into SysML
Properties of a Structured Requirement-stereotyped element (Fig. \ref{fig:Stereotypes}). Thus corresponding model elements
and even diagrams may slot-in as model-based supplements to the textual ``shall'' statement in alignment with the Information-based Needs and Requirements Definition and Management meta-model adopted by the GtWR \citep{wheatcraft2019INRDM,NRM2022} (Fig. \ref{fig:RelationMap}).
This meta-modeling method is well-supported by SysML profiles and the \citet{SysML1.7} specification
 which describes how SysML requirements can be extended to define requirement types and Property-Based Requirements (Sec. \ref{sec:MathBased}). MBSR goes ``beyond treating a requirement expression as an indivisible entity and allows the terms inside the requirement statement to be referenced'' \citep{dick2017integrating} as \SA{commonly done in SysML modeling tools}.

%%%%
%%%%

%%%%
%%%%

Earlier meta-models of MBSR \citep{herber2022model} limited the attribute types to
corresponding SysML types such as Block for the Subject pattern slot, but this
approach was later found to be too restrictive and the standard SysML
meta-model decision to use NamedElement was adopted \citep{herber2023}.
The MBSR meta-model works by using Generalization relationships with the
existing SysML Requirement, taking advantage of the standard syntax and
semantics of SysML Requirements while separating concerns of the
Structured Requirement pattern slots, and an organization's conventional requirement attributes.
%<*mytagR1C8-1>
Organization attributes are customizable and might
include a secondary \SA{system} V\&V method, unique identifier of the associated Work Breakdown Structure unit,
or a `short text' which provides an informal requirement statement in layman's terms.
Other organization attributes may be required for compliance, for syncing with the RMT in use, and for ontological coherence with the ASoT.
%</mytagR1C8-1>

%%%%%%%%%%%%%%%%%%%%%%%%%%%%%%%%%%%%%%%%%%%%%%%%%%%%%%

\section{INCOSE-Derived MBSR for Digital Requirements Engineering}
\label{sec:INCOSE-MBSR}

The MBSR extensions presented in this paper (\SA{abridged in} Fig. \ref{fig:Stereotypes}) explore the utility
of adding the 49 Attributes\SA{,} 42 Rules\SA{, and 15 Characteristics} described in the INCOSE GtWR
with appropriate SysML standard and custom types defined.
\SA{Extending the prior MBSR profile (Section \ref{sec:MBSR}) in this way provided readily-usable SysML classifiers compatible with the INCOSE GtWR ontology, and embedded specific guidance into the SysML modeling environment to aid requirements development and requirements V\&V.}

\subsection{Meta-Model of Rules, Attributes, Characteristics}
\label{sec:GtWR-Metamodel}

%%%%
\begin{figure}[bt]
\centering
\includegraphics[width=\columnwidth]{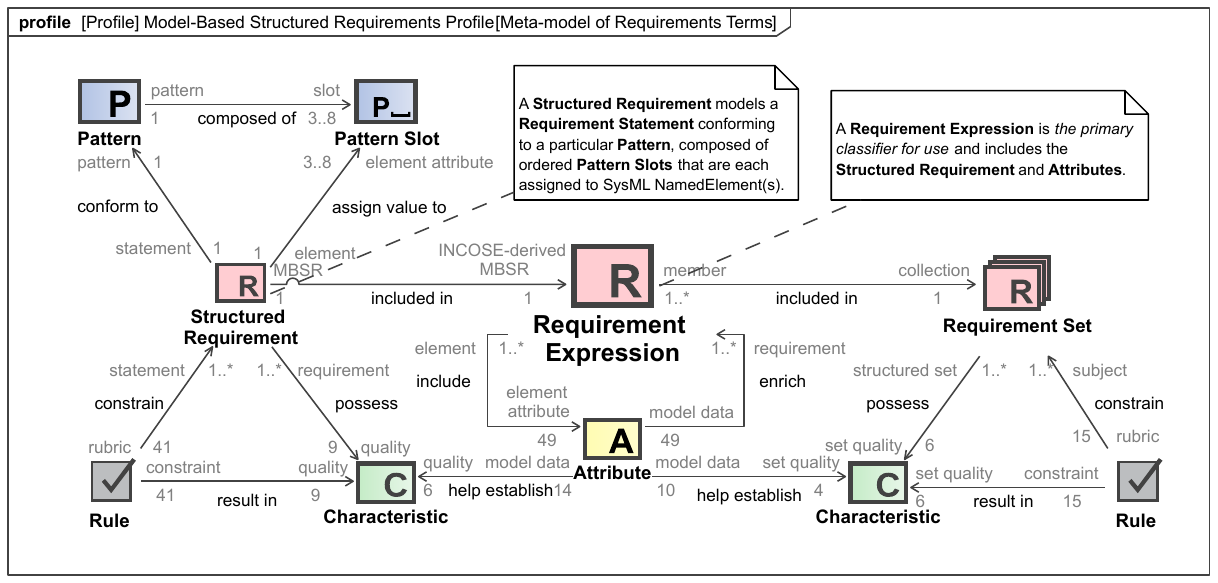}
\caption{SysML meta-model of INCOSE-derived MBSR classifiers derived from Figure 4 of \citet{GtWR2023}.}
\label{fig:Metamodel}
\end{figure}
%%%%

Figure \ref{fig:Metamodel} \SA{presents} a \SA{SysML} meta-model \SA{derived from the Entity-Relationship Diagram in Figure 4 of} GtWR \SA{(\citet{GtWR2023})} that clearly relates these DRE terms to each other while respecting their given definitions. The ``Requirement Statement'' term is replaced with the MBSR ``Structured Requirement'' and \SA{additionally} relates Patterns and Pattern Slots; and the ``Requirement Expression'' is emphasized as \textit{the primary \SA{classifier} for use} so as not to confuse it with the standard SysML Requirement (Sec. \ref{sec:ClassicalSysML}) in the tool\SA{'s user} interface. A ``Requirement Set'' contains Requirement Expressions and potentially Requirement Sets, and is a subclass of Requirement Expression rather than a SysML Package to maintain uniform application of the GtWR ontology in the SysML profile.
UML-based multiplicity and roles were used on directed associations \SA{in Fig. \ref{fig:Metamodel}} to reflect the cross-reference matrices in the GtWR.

A Requirement Set (historically called requirement modules or composite requirements)
is defined to distinguish from an individual Requirement Expression as in the GtWR, and provides
additional querying, filtering, and meta-modeling capabilities for
isolating Sets. Likewise, Needs and Need Sets are defined and inherit
some of the same Attributes, with a different icon to visually
distinguish them from requirements, potentially addressing a major concern with standard SysML voiced in the GtWR.

%<*mytagR3C10-2>
Attributes help establish Characteristics, of which there are exactly 9
for Needs and Requirements, and 6 for Need Sets and Requirement Sets
(\SA{see} Table \ref{tab:Characteristics} \SA{in Sec. \ref{sec:ApplicabilityToStandardGuidance} for details}).
%</mytagR3C10-2>
%<*mytagR4C6>
The Attributes are grouped according to their purpose and numbered to maintain order
and to aid in searching; some Attributes are marked with an ending asterisk (*) to indicate membership
in the minimum set of Attributes according to the GtWR. Notably, some
Attributes are already defined elsewhere, such as A15 (Unique Identifier) and A16
(Unique Name), and are modeled using \Stereotype{Customization} derived properties that query the
standard SysML properties for completeness of the INCOSE-derived MBSR profile \SB{(Fig.~\ref{fig:Customizations})}.
%</mytagR4C6>

\begin{figure}[p]
\includegraphics[width=\columnwidth]{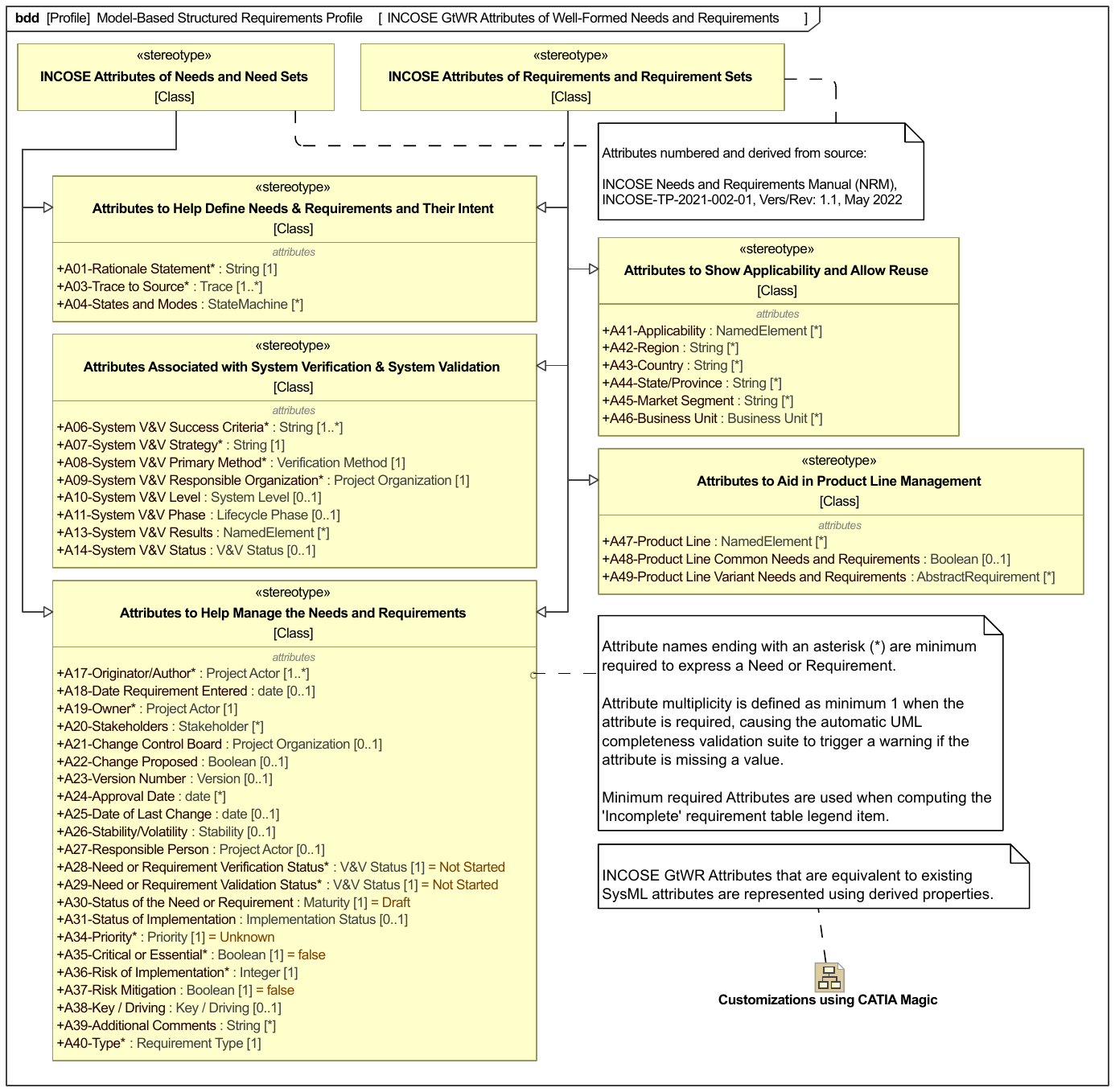}
\caption{Model-based INCOSE GtWR Attributes of Well-Formed Needs and Requirements. Refer to Fig.~\ref{fig:Stereotypes} for further relationship definitions\SB{, and Fig.~\ref{fig:Customizations} for customizations.}}
\label{fig:GtWR-Attributes}
\end{figure}

Value Types such as Enumerations (e.g. Verification Method) and organization-related Stereotypes (e.g. Project Actor, Business Unit) with the Class meta-class were created and selected based on the Attribute definitions and guidance in the NRM. These Value Types and set of minimum viable Attributes should be harmonized with the organization's conventions, e.g. requirement types, statuses, and risks (Fig. \ref{fig:GtWR-Attributes}).
%<*mytagR1C8-2>
\SA{The Organization Requirement Attributes shown in Fig. \ref{fig:GtWR-Attributes} are representative, and their modification and use depends on the organization's context and conventions. This list of attributes should conform to the Organizational Requirement Definition Requirements, as represented in \citet{GtWR2023} Figure 7. The presence of such attributes here highlights the flexibility of the MSBR technique while addressing potential concerns that some attributes common in the organization, such as a secondary system V\&V method, are missing from the GtWR attributes.}
%</mytagR1C8-2>

%%%%
%%%%

\subsection{Requirements V\&V using the Profile}
\label{sec:RequirementsVandV}

%<*mytagR1C7>
\SA{We emphasize the distinction between requirements V\&V and system V\&V, and until the release of the INCOSE Requirements Working Group technical products \citep{GtWR2023,NRM2022}, the common systems engineering parlance has been that ``requirements verification'' refers to system verification against the requirements. Requirements V\&V refers to the verification and validation of the requirement expressions themselves, and here we present requirements V\&V using the SysML modeling tool and INCOSE-derived MBSR SysML profile.}%
%</mytagR1C7>

%<*mytagR1C11>
A metric suite is provided in the MBSR Profile to demonstrate how \SA{SysML-}validation-based metric definitions
and custom scripting can be used to compute completeness metrics on an MBSR set \citep{herber2022model,herber2024repo}.
The metric suite works by referencing \SA{SysML} validation rules that check if each MBSR Pattern Slot is filled.
%</mytagR1C11>
The metric table using it shows how many MBSRs are in a given Package at a certain date and time, how many set a value for each Pattern Slot,
and what percentage of the MBSRs complete the requirement Pattern. The intended usage is as follows:
\begin{enumerate}
 \item Create a new metric table under a suitably named Package.
 \item Set the ``Metric Suite'' under ``Criteria'' to ``MBSR Completeness'' from the MBSR Profile.
 \item Click ``Calculate Metrics'' and select ``Add New Metric with Different Parameters''.
 \item Ensure that ``Scope'' and ``Type'' columns are shown by selecting them from the ``Columns'' dropdown menu.
 \item Set the Scope value of the new metric instance to the Package containing MBSRs.
 \item Set the Type value to the ``Structured Requirement'' from the MBSR Profile.
 \item Click ``Calculate Metrics'' and select ``Recalculate''.
 \item Export the metric table to CSV or XLSX for downstream workflows, or
 \item Query the metric table elements from report templates to generate custom reports.
\end{enumerate}

The MBSR Completeness metric suite may be used in concert with traceability metric suites to create requirement metrics dashboards. Computing metrics in this way can be more effective than checking every requirement in a matrix, and it provides timestamped data that may be used for burndown charts or compliance audits. See Sec. \ref{sec:FutureWork} for a discussion of future work on MBSR metrics.

\begin{figure}[h]
\centering
\includegraphics[width=0.5\columnwidth]{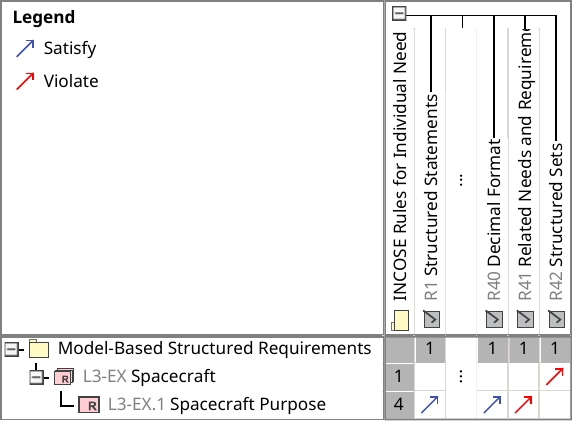}
\caption{A Requirements Satisfaction Matrix of INCOSE GtWR Rules (abbreviated), with Violate relationships shown in red.}
\label{fig:Matrix}
\end{figure}

%<*mytagR3C17-2>
When a Requirement Expression or Requirement Set is verified and
validated, a SysML \Stereotype{satisfy} relationship is added from each requirement
to the respective Characteristic, providing model data and metrics of their well-formedness.
Like Attributes, Rules help establish Characteristics of a well-formed Requirement Expression or Requirement
Set, and during requirements V\&V activities, a \Stereotype{satisfy}
or new \Stereotype{Violate} relationship is likewise created for metrics and feedback (Fig. \ref{fig:Matrix}).
The Rules and Characteristics linked to each requirement may be shown in requirement tables\SA{, metric and generic tables,} and \SA{rendered} in custom reports.
%</mytagR3C17-2>

%%%%
%%%%

%%%%%%%%%%%%%%%%%%%%

\subsection{Application and Case Study}
\label{sec:ApplicationCaseStudy}

\begin{figure}[bt]
\centering
\includegraphics[width=\columnwidth]{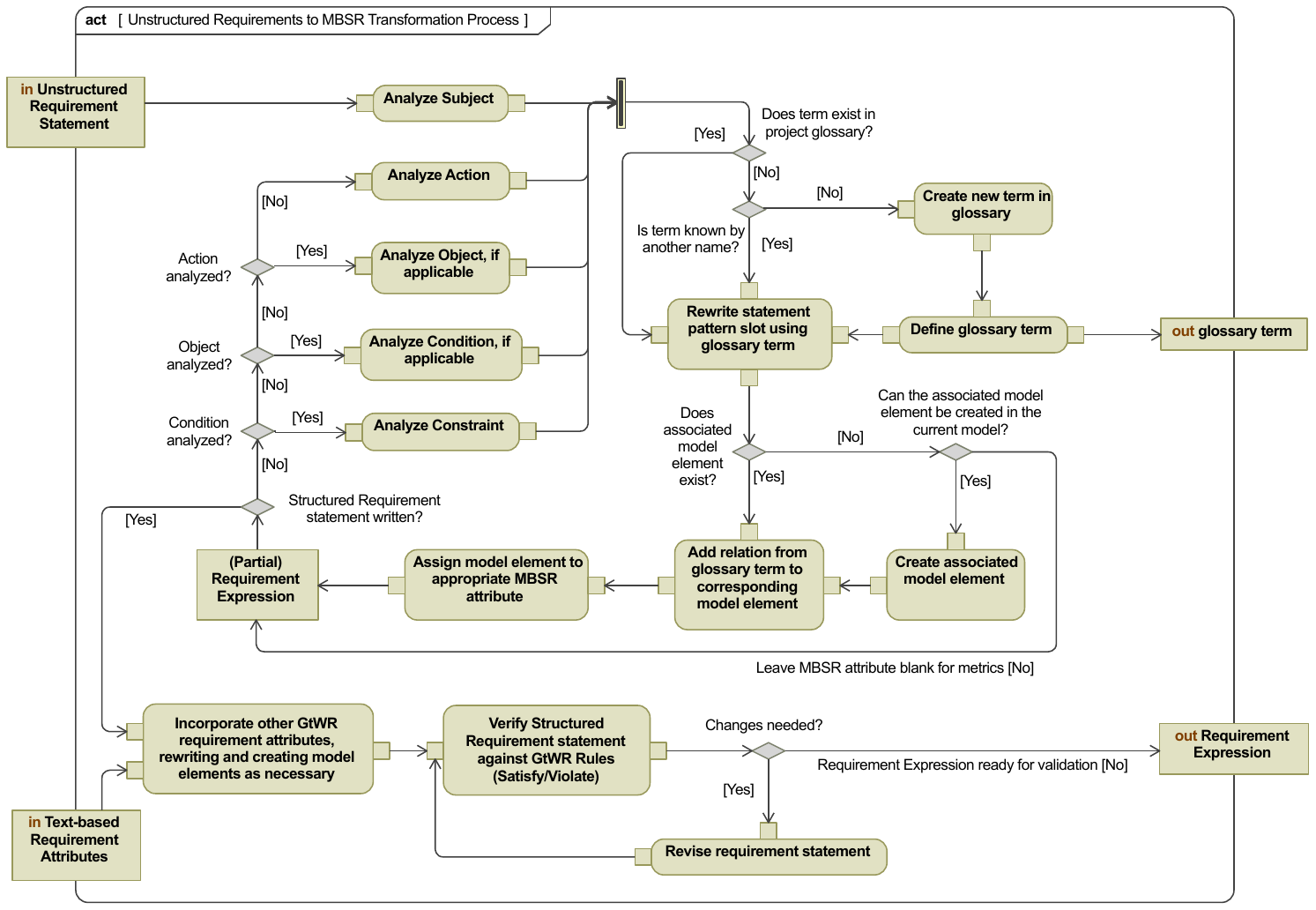}
\caption{\SB{Activity diagram of the transformation of unstructured requirement and text-based attributes to verified INCOSE-derived MBSR requirement expression.}}
\label{fig:TransformationProcess}
\end{figure}

\SA{Having presented the INCOSE-derived MBSR profile above (Sec.~\ref{sec:INCOSE-MBSR}), we now show how the SysML profile and technique are applied using an example in the space domain. The Mars Sample Return project case study is also discussed briefly to explain how the INCOSE-derived MBSR profile was used in a real-world project, starting with unverified draft requirements. Whether the requirements engineer is converting requirement statements to MBSRs or they are developing new requirements in tandem with the system architecture modeling activities, the process is much the same \SB{(Fig.~\ref{fig:TransformationProcess})}.}

%%%%
%%%%

%<*mytagR3C12>
%<*mytagR0C4>
\SA{The INCOSE-derived} MBSR SysML profile \citep{herber2024repo}
was used to develop over 300 requirements and 50 requirement sets
at NASA Jet Propulsion Laboratory (JPL) for the Mars Returned Sample Handling
project in Pre-Phase A \SA{\citep{Wheaton2025a,Younse2025}}.
%</mytagR0C4>
%<*mytagR5C33>
Drafts of requirements were received from the subsystem engineering leads \SA{in a spreadsheet},
\SA{and} imported into No Magic Cameo Systems Modeler 2022x \citep{Cameo2022x} \SA{using the requirements table diagram and the Excel import feature.}
\SA{The} given information, such as rationale, verification method, verification approach, and additional
comments \SA{were first inserted into their} respective GtWR-derived Attributes\SA{, and continued to be revised during requirements development}.
%</mytagR5C33>
%</mytagR3C12>
%<*mytagR3C17>
Matrix diagrams were created, as in Fig. \ref{fig:Matrix}, to aid in the revision of the requirement
statements, providing visual feedback as a kind of checklist, and data
for metrics used to triage requirements for later revision. All Rules,
Attributes, and Characteristics contained the corresponding documentation
from GtWR to further assist the usage of the Profile with tooltips.
%</mytagR3C17>
To model Defined Terms as described in GtWR, Cameo Systems Modeler Glossary
Tables were filled with Terms and synonyms---often acronyms---active
hyperlinks to the definition source, and SysML Allocate relationships to
system model elements and diagrams, providing automatic underlining of
Defined Terms used consistently throughout requirements statements. In
addition, other meta-model elements were added to the Profile to capture
the broader range of related information: Goal, Assumption, Project
Actor / Role / Organization, Requirement Types relevant to NASA / JPL.

\begin{figure}[p]
\centering
\includegraphics[width=\columnwidth]{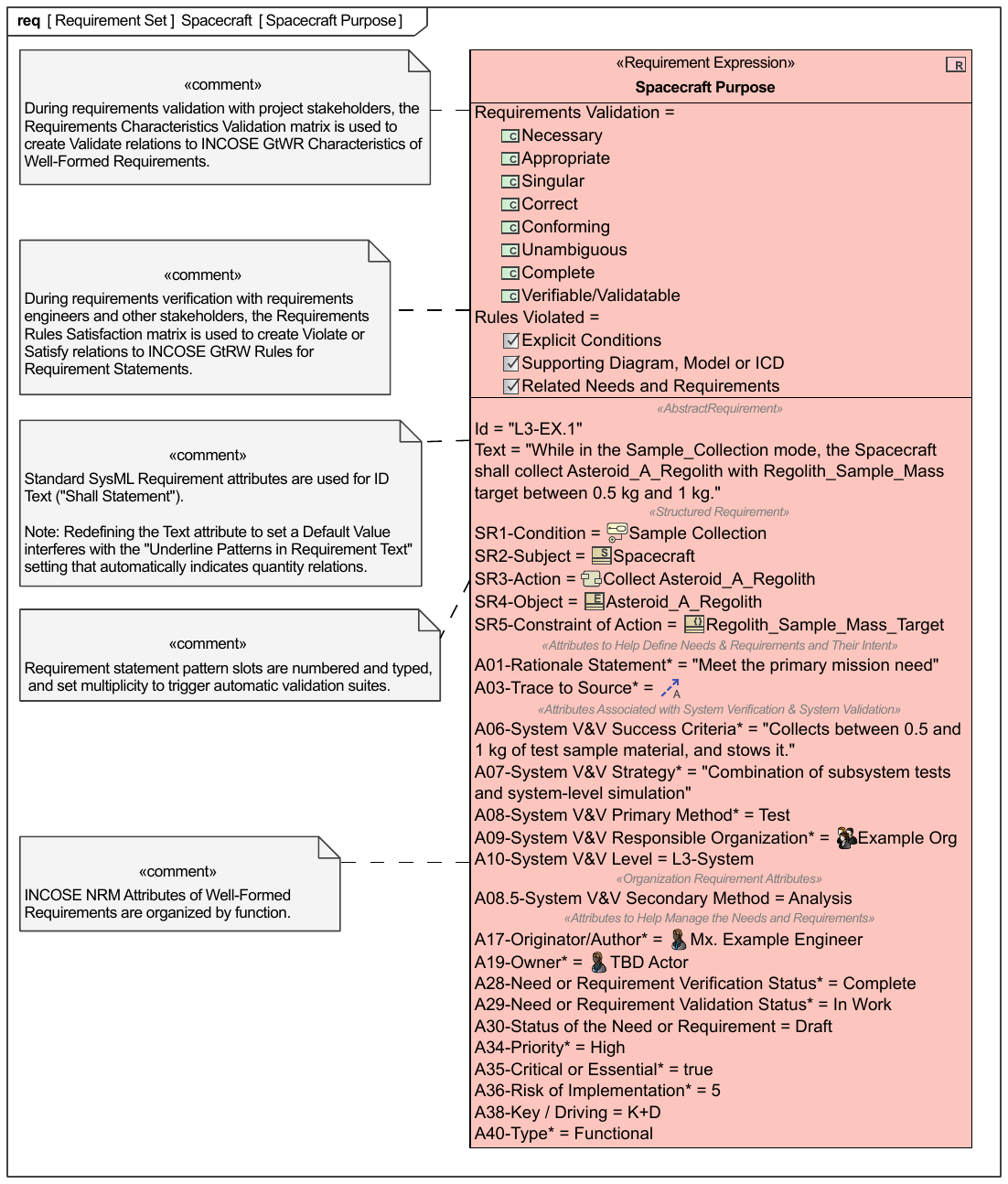}
\caption{Single example MBSR with INCOSE GtWR Attributes, Characteristics, and Rules.}
\label{fig:Example}
\end{figure}

\SA{In Fig.~\ref{fig:Example}, we show how a singular INCOSE-derived MBSR may be rendered in a SysML requirements diagram. By showing and hiding compartments and attributes of interest, a custom view is rendered to meet stakeholder concerns. At a glance, a particular requirement expression is shown to be validated according to individual characteristics (Fig. \ref{fig:Metamodel}), and remaining rules in violation may be shown to draw attention while reviewing the requirement statement. The classical SysML requirement attributes of Id and Text are shown first, followed by categorized attributes numbered accordingly. Lastly, the base MBSR pattern slots are rendered with their associated values' icons to emphasize the model-based nature of the view. Figure \ref{fig:XMI} shows much of the same information as Fig. \ref{fig:Example} but using the standard XML-based data interchange format in support of the tool diversity present in digital engineering environments.}

%%%%
%%%%

%%%%
\begin{figure}[h]
\lstset{language=XML,
  basicstyle=\ttfamily,
  columns=fullflexible,
  showstringspaces=false,
  morestring=[s]{'}{'},
  stringstyle=\color{darkgreen},
  morekeywords={xmi:id,base_Class,Id,Text,SR1_Condition,SR2_Subject,SR3_Action,SR4_Object,SR5_Constraint_of_Action,A01_Rationale_Statement_,A08_System_V_V_Primary_Method_,A10_System_V_V_Level,A28_Need_or_Requirement_Verification_Status_,A30_Status_of_the_Need_or_Requirement,A34_Priority_,A38_Key___Driving,A40_Type_}
}
\centering
\begin{lrbox}{\mybox}%
\begin{lstlisting}
<Model_Based_Structured_Requirements_Profile:Requirement_Expression
xmi:id='_2022x_2_46d01c0_1707158979643_806727_21479_'
base_Class='_2022x_2_46d01c0_1707158979643_806727_21479'
Id='L3-EX.1'
Text='While in the Sample Collection mode, the Spacecraft
      shall collect Asteroid A Regolith with Regolith Sample Mass
      target between 0.5 kg and 1 kg.'
SR1_Condition='_2022x_2_46d01c0_1707158979575_932108_21295'
SR2_Subject='_2022x_2_46d01c0_1707158979604_996809_21373'
SR3_Action='_2022x_2_46d01c0_1707158979574_907885_21293'
SR4_Object='_2022x_2_46d01c0_1707158979726_396359_21900'
SR5_Constraint_of_Action='_2022x_2_46d01c0_1707158979579_344791_21322'
A01_Rationale_Statement_='Meet the primary mission need'
A08_System_V_V_Primary_Method_='Test'
A10_System_V_V_Level='L3-System'
A28_Need_or_Requirement_Verification_Status_='Complete'
A30_Status_of_the_Need_or_Requirement='Draft'
A34_Priority_='High'
A38_Key___Driving='K+D'
A40_Type_='Functional' />
\end{lstlisting}%
\end{lrbox}%
\scalebox{0.65}{\usebox{\mybox}}
\caption{UML 2.5 XMI definition of the example MBSR shown in Fig. \ref{fig:Example} (newlines added \& attributes reordered for readability).}
\label{fig:XMI}
\end{figure}
%%%%

\begin{figure}[bt]
\centering
\includegraphics[width=\columnwidth]{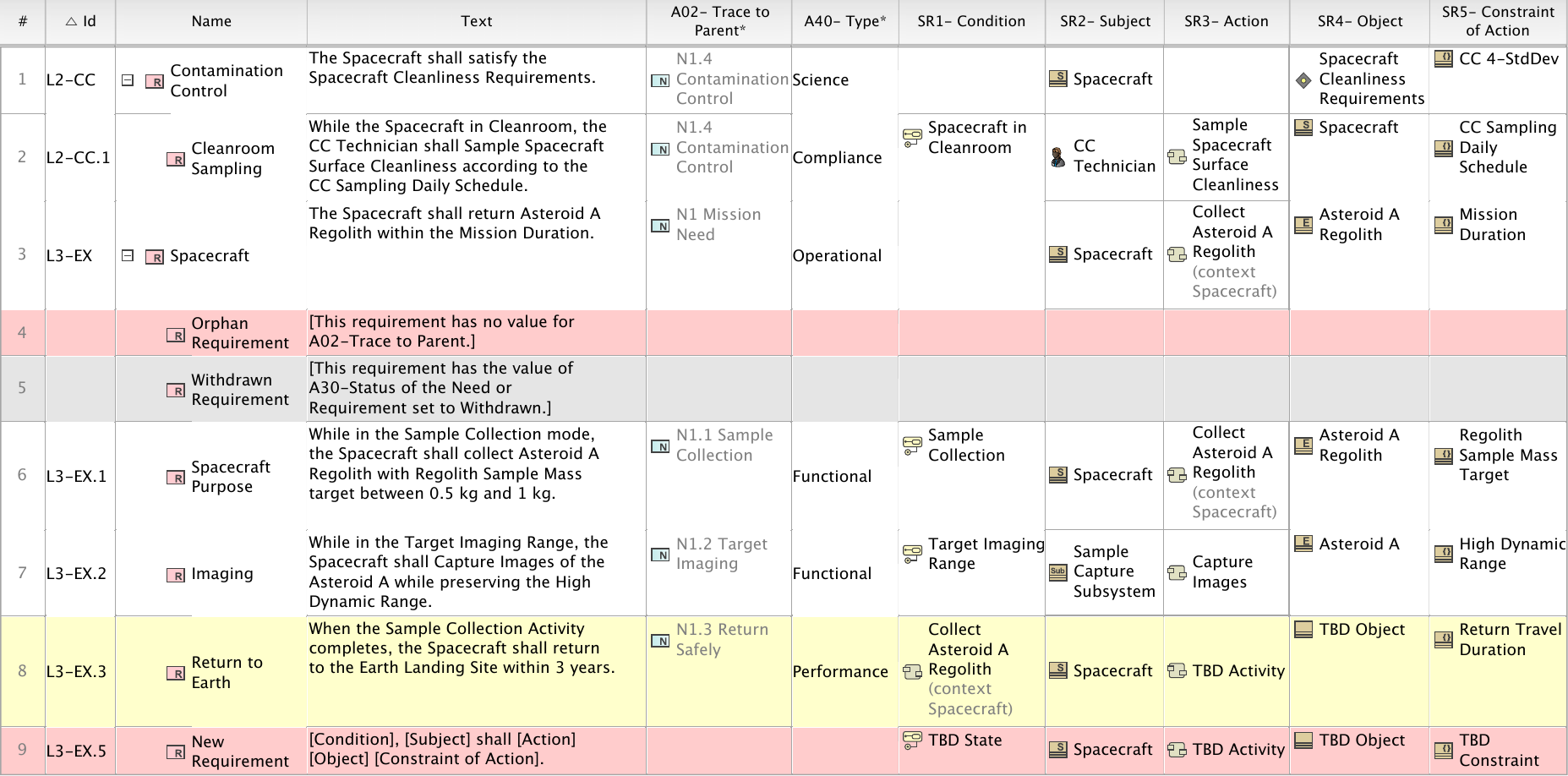}
\caption{INCOSE-derived MBSRs in a requirement table with model-based legend highlighting.}
\label{fig:ReqsTable}
\end{figure}

%<*mytagR1C9>
\SA{While Fig. \ref{fig:Example} is shown here more for demonstration purposes, the requirement table is a more common view due to its compactness and ability to sort columns. Figure \ref{fig:ReqsTable} shows how INCOSE-derived MBSRs may be viewed as a nested tree in a requirements table, with an icon accompanying each model element to again emphasize its model-based nature. Here we also show the numbered characteristics validated, and the numbered rules violated to assist in requirements V\&V activities. All structured requirement pattern slots are shown, with some being filled in with ``TBD'' elements to clearly identify parts of the architecture yet to be modeled and which may be easily searched and enumerated in a separate ``TB(X)'' table. Attribute ``A02-Trace to Parent'' is also shown to emphasize traceability of top-level requirements to need expressions in the model. Lastly, a model-based legend is used to clearly identify orphan requirements (red), withdrawn (but not deleted) requirements (gray), and incomplete requirements (yellow). Complete requirements according to their pattern slots are shown in the table with a clear or white background. Such a legend is easily modified using structured expressions and scripts, and, for example, may only show requirements as complete when their minimum attribute set has been filled.}
%</mytagR1C9>

%%%%
%%%%

%%%%%%%%%%%%%%%%%%%%%%%%%%%%%%%%%%%%%%%%%%%%%%%%%%%%%%

\section{Discussion}
\label{sec:Discussion}

%<*mytagR2C6>
\SA{INCOSE-derived MBSR is a SysML language extension which is leveraged during typical requirements engineering activities, and which is compatible with existing RMT integration. However, compared to an RMT, which supports additional requirement attributes, MBSR validates not only standard data types such as dates and numbers, but also any custom elements created in an architecture model and its accompanying profile. Compared to RMTs, which would require similar customization to include the GtWR rules and characteristics, the MBSR-enabled SysML model makes requirements V\&V data directly accessible, supporting a digital engineering approach of data-driven decision-making in a variety of stakeholder views.}
%</mytagR2C6>

The immediate advantage of this MBSR approach was the ability to keep
architecture modeling activities confined to a single tool, and without
the loss of expressiveness that would normally result from using classical SysML Requirements.
Report templates were created using the full MBSR expression, and according to
stakeholder expectations for engineering reviews.
%<*mytagR5C28>
The system architecture model and cohabitated requirements were exported
to targeted stakeholder views, including custom PowerPoint templates for
an architecture overview and a Requirement Set formal review,
an internal website (Web Report), and Excel spreadsheets (adding
relevant table columns as needed).
%</mytagR5C28>
%<*mytagR3C19>
While this targeted approach of INCOSE-derived MBSR usage received overwhelmingly positive feedback from the team, it came with its own challenges that became evident as limitations of the tool were encountered \SA{(Sec. \ref{sec:Challenges})}.
%</mytagR3C19>

\subsection{Observed Benefits}
\label{sec:Benefits}

%<*mytagR3C15>
Observed benefits \SA{compared to unstructured document-based and classical SysML requirements} during the {NASA}~\slash~{ESA}~\slash~{JPL} joint Mars Sample Return project included: improved requirement statement quality due to the use of a common requirement Pattern, Rules, and accompanying Attributes; increased accuracy and assurance of model views using generated custom reports; access to system model elements for increased traceability and specificity; and model-assisted traceability views of Key Driving Requirements across multiple levels of the system hierarchy. Table \ref{tab:Benefits} provides a more complete list with perceived impact on the project \SA{during Pre-Phase A}.
%</mytagR3C15>

%%%%
\begin{table}[p]
\footnotesize
\caption{Benefits and perceived impacts of \SA{using} INCOSE-derived MBSR \SA{for DRE}. Scale: \highimpact~High, \medimpact~Medium, \lowimpact~Low.}
\label{tab:Benefits}
\begin{tabular}{p{0.8\columnwidth}c}
\textbf{Observed Benefit} & \textbf{Impact}\\
\hline%
Access to the full system architecture model supports creation of the ASoT with rich traceability & \highimpact \\
\hline%
Combined use of a singular tool is cost-effective & \highimpact \\
\hline%
Consistent use of requirement statement patterns aids V\&V of system elements & \highimpact \\
\hline%
Glossary terms are underlined throughout the model, with multiple definitions visible in the tooltip & \highimpact\\
\hline%
Requirements are exportable in custom Word / Excel / PowerPoint model-based reports & \highimpact\\
\hline%
SysML meta-modeling capabilities support extensive customization & \highimpact \\
\hline%
Full ISO 80000 units of measure and MARTE real-time SysML profiles are available to extend and use & \highimpact\\
\hline%
TBX summary table is easily made with scope, and TB{[}CDRN{]} regular expression as a filter on all fields & \highimpact\\
\hline%
Custom-query Relation Map diagrams expose MBSR relationships tailored to stakeholder concerns & \medimpact\\
\hline%
Requirement IDs are customizable, with predictable increments, and may be typed-in directly in diagrams & \medimpact\\
\hline%
Collaboration plugin supports simultaneous team use with change control and precise feedback capability & \medimpact \\
\hline%
Any Attribute is displayable and sortable in a table & \medimpact\\
\hline%
Matrices with embedded tooltip documentation and double-click entries supports efficient workflows & \medimpact\\
\hline%
Custom legend items adorn tables and diagrams using custom scripts or Structured Expressions & \medimpact\\
\hline%
Requirement tables may be imported/synchronized with Excel spreadsheets or use data connectors & \medimpact\\
\hline%
MBSR Attributes are quickly searchable/filterable by number, e.g. ``A28'' or ``SR'' & \medimpact\\
\hline%
Only used Attributes appear on symbols by default & \medimpact\\
\hline%
Derived Properties defined by custom scripts or Structured Expressions enhance model-based definition & \medimpact\\
\hline%
UML 2.5 XMI exports are complete with model element reference identifiers & \lowimpact\\
\hline%
Copy/paste of values into multiple table cells simultaneously works in the simple case & \lowimpact\\
\hline% Please only put a hline at the end of the table
\end{tabular}
\end{table}
%%%%

\subsection{Observed Challenges}
\label{sec:Challenges}

Applying INCOSE-derived MBSR to a real project met with real challenges \SA{(Table \ref{tab:Challenges})}, primarily that the SysML tool in use \SA{was} slow to generate reports, render views of model data and adorning legend items, and save model updates, causing significantly delayed cycle times during MBSR meta-modeling and modeling of the system. Other challenges of this approach relate to the SysML tool's lack of RMT facilities that would automate some model updates and provide high assurance of their correctness. In this case, only one engineer was developing the requirements at the time, and formal change management had not been activated during this technology development phase. A more detailed list of the observed challenges and perceived impacts on the project are sorted in Table \ref{tab:Challenges}.

%%%%
\begin{table}[p]
\footnotesize
\caption{Challenges and perceived impacts of \SA{using} INCOSE-derived MBSR \SA{for DRE}. Scale: {\highimpact}~High, {\medimpact}~Medium, {\lowimpact}~Low.}
\label{tab:Challenges}
\begin{tabular}{p{0.8\columnwidth}c}
\textbf{Observed Challenge} & \textbf{Impact}\\
\hline%
Careless mistakes made in the Shared Profile may destroy system model information & {\highimpact}\\
\hline%
Custom report templates are time-consuming and error-prone to create & {\highimpact}\\
\hline%
Reports are slow to generate, increasing cycle times & {\highimpact}\\
\hline%
Writing custom scripts is challenging due to often inadequate documentation & {\highimpact}\\
\hline%
Requirements management Attributes such as `Date of Last Change' and `Version Number' require manual entry workflows & {\medimpact}\\
\hline%
Newly filled Attributes will appear in all affected diagrams, mangling diagrams unless compartments are manually suppressed & {\medimpact}\\
\hline%
Glossary terms sometimes do not underline, are never underlined in the Web Report, or the tooltip does not reliably appear & {\medimpact}\\
\hline%
Redefining \Stereotype{AbstractRequirement} Text attribute breaks automatic underlining of quantity relations & {\medimpact}\\
\hline%
Table adorning / loading can be slow & {\medimpact}\\
\hline%
Legends cannot adorn table cells, only table rows & {\medimpact}\\
\hline%
ReqIF exports require manual attribute mapping, and do not include referenced system model elements & {\medimpact}\\
\hline%
Requirement IDs are tedious to set/increase/decrease with Element Numbering dialogs & {\medimpact}\\
\hline%
Attributes with multiplicity >1 fails to copy/paste; paste chooses the first element with that name, not the same exact element copied & {\medimpact}\\
\hline%
\SA{Element naming conflicts interfere with the use of copy/paste in tables} & {\medimpact}\\
\hline%
Requirement IDs may conflict and can lose their order & {\lowimpact}\\
\hline%
Attributes appear under multiple groups in table column selection dialog & {\lowimpact}\\
\hline%
SysML Properties may only have one (1) Owner, preventing reuse in organization-defined Attribute Sets & {\lowimpact}\\
\hline%
Table scrolling performs sequential loading, temporarily revealing blank rows in large requirement tables & {\lowimpact}\\
\hline%
Glossary term allocations to model elements may be duplicative of MBSR attribute values & {\lowimpact}\\
\hline% Please only put a hline at the end of the table
\end{tabular}
\end{table}
%%%%

\subsection{Applicability to Standard Guidance}
\label{sec:ApplicabilityToStandardGuidance}

%<*mytagR3C14>
This INCOSE-derived MBSR Profile supports the requirements engineering practices according to
the NASA SE Handbook guidelines and Appendix C checklist \citep{NASA_SE_HB2017}.
Although the NASA SE Handbook does not present clearly defined characteristics, the glossary entry and Appendix C checklist contains language that may be mapped to all of the INCOSE GtWR Characteristics (refer to Table \ref{tab:Characteristics}): 
(C1) `necessary \dots to meet mission and system goals and objectives';
(C2) `compliance';
(C3) `unambiguous in meaning' and `complies with the project's template and style rules';
(C4) `completeness';
(C5) `not redundant';
(C6) `feasible to obtain';
(C7) `verifiability/testability';
(C8) `correct';
(C9) `clear' and `clarity';
(C10) `completeness';
(C11) `consistency' and `not in conflict with one another';
(C12) `technically feasible';
(C13) `clarity' and `adequately related with respect to terms used';
(C14) `can be validated'; and
(C15) `correctness' and `completeness'.
By contrast, the \citet{ISO29148} standard provides definitions for 14 out of the 15 Characteristics and clearly categorizes them for individual requirements and for requirement sets.
%</mytagR3C14>
MBSR supports the bidirectional traceability and the creation
of NASA-requested artifacts such as the Requirements Allocation Sheet,
TBX report, and Requirements Verification and Validation Matrices.
Further discussion of the relationships among the NASA SE Handbook, the ISO 29148:2018 standard, and the INCOSE GtWR
may be found in the ontology section of the NRM.
While the NASA SE Handbook and INCOSE GtWR (and related guides and manuals) are complementary,
the GtWR provided enumerated and defined precision amenable to SysML
meta-modeling and reuse in the system architecture model.

%%%%
\begin{table*}[bt]
\caption{INCOSE Characteristics of well-formed sets and individual needs and requirements \citep{GtWR2023} with mapping to NASA and ISO guidance \citep{NASA_SE_HB2017,ISO29148}.}
\label{tab:Characteristics}
\begin{tabular}{p{0.03\linewidth}p{0.18\linewidth}p{0.28\linewidth}p{0.23\linewidth}cc}
\textbf{ID} & \textbf{Name} & \textbf{Applicability} & \textbf{Derivation} & \textbf{NASA?} & \textbf{ISO?}\\
\hline%
C1 & Necessary & Needs \& Requirements & Formal Transformation & $\CheckmarkBold$ & $\CheckmarkBold$\\
\hline%
C2 & Appropriate & Needs \& Requirements & Formal Transformation & $\CheckmarkBold$ & $\CheckmarkBold$\\
\hline%
C3 & Unambiguous & Needs \& Requirements & Agreed-to Obligation & $\CheckmarkBold$ & $\CheckmarkBold$\\%`clear' \& `unambiguous in meaning'\\
\hline%
C4 & Complete & Needs \& Requirements & Agreed-to Obligation & $\CheckmarkBold$ & $\CheckmarkBold$ \\
\hline%
C5 & Singular & Needs \& Requirements & Formal Transformation & $\CheckmarkBold$ & $\CheckmarkBold$\\%`not redundant'\\
\hline%
C6 & Feasible & Needs \& Requirements & Agreed-to Obligation & $\CheckmarkBold$ & $\CheckmarkBold$\\%`feasible to obtain'\\
\hline%
C7 & Verifiable & Needs \& Requirements & Agreed-to Obligation & $\CheckmarkBold$  & $\CheckmarkBold$\\
\hline%
C8 & Correct & Needs \& Requirements & Formal Transformation & $\CheckmarkBold$ & $\CheckmarkBold$\\%`correct'\\
\hline%
C9 & Conforming & Needs \& Requirements & Formal Transformation & $\CheckmarkBold$ & $\CheckmarkBold$\\%`clear'\\
\hline%
C10 & Complete & Need Sets \& Requirement Sets & Formal Transformation & $\CheckmarkBold$ & $\CheckmarkBold$\\%`not in conflict with one another'\\
\hline%
C11 & Consistent & Need Sets \& Requirement Sets & Formal Transformation & $\CheckmarkBold$ & $\CheckmarkBold$\\%`not in conflict with one another'\\
\hline%
C12 & Feasible & Need Sets \& Requirement Sets & Agreed-to Obligation & $\CheckmarkBold$ & $\CheckmarkBold$\\%`feasible to obtain'\\
\hline%
C13 & Comprehensible & Need Sets \& Requirement Sets & Agreed-to Obligation & $\CheckmarkBold$ & $\CheckmarkBold$\\%`adequately related with respect to terms used'\\
\hline%
C14 & Able to be validated & Need Sets \& Requirement Sets & Agreed-to Obligation & $\CheckmarkBold$ & $\CheckmarkBold$\\%`can be validated'\\
\hline%
C15 & Correct & Need Sets \& Requirement Sets & Formal Transformation & $\CheckmarkBold$ &\\
\hline  % Please only put a hline at the end of the table
\end{tabular}
\end{table*}
%%%%

%%%%%%%%%%%%%%%%%%%%%%%%%%%%%%%%%%%%%%%%%%%%%%%%%%%%%%

\section{Related Work}
\label{sec:RelatedWork}

\citet{salado2023model} presents three approaches to model-based requirements found in the literature: 1) dedicated classes and flagged models, 2) math- and property-based models of requirements, and 3) semantic extensions to model the problem space. The relationship of MBSR to template-based textual requirements is discussed in Sec. \ref{sec:StructuredReq}\SA{,} and \SA{classical SysML requirements} (``dedicated classes'') are discussed in Sec. \ref{sec:SysML}. This section provides a brief overview of related model-based \SA{requirement} approaches
%<*mytagR4C3-1>
while noting similarities and differences with MBSR as presented in this paper.
%</mytagR4C3-1>

\subsection{System Models as Requirements}
\label{sec:FlaggedModels}

\citet{wach2022need} presents a model-based method for capturing requirements using standard SysML elements and diagrams other than the textual-based SysML \Stereotype{AbstractRequirement}. As discussed in their paper, this method unnecessarily constrains the solution space and is therefore considered poor requirements engineering practice. Flagged models as requirements will be unfamiliar to stakeholders and may cause significant confusion compared to ``shall'' statements \citep{NRM2022}.
%<*mytagR4C3-2>
MBSR extends the use of textual statements with deeper connectivity to the system model compared to standard SysML, without relying on SysML diagrams to model the problem space. Statement patterns are used in MBSR both to assist readability and to facilitate model definition and traceability. While SysML elements are used in MBSR slots, they may remain placeholders until design activities commence and more details of the solution space are known.
%</mytagR4C3-2>

\subsection{Mathematical Models of Requirements}
\label{sec:MathBased}

Wymorian theory of requirements is based in set theory \citep{wymore1993model} and is not directly applicable to SysML models. This formulation of the problem space enables mathematical queries to assist in requirements V\&V and may in particular assess completeness of a requirement set. The basic modeling construct is called a system design requirement and is defined as a sextuple including: 1) input/output requirement, 2) technology requirement, 3) performance requirement, 4) cost requirement, 5) trade-off requirement, and 6) system test requirement.
%<*mytagR4C3-3>
Compared to Wymorian theory of requirements, MBSR is more relevant to modern MBSE practice because of its integration with SysML models. Integration of this alternative MBSE theory with SysML remains an open area of research \citep{wach2019can}, and SysML v2 may provide new research opportunities due to its strong semantic foundations compared to v1.
%</mytagR4C3-3>

Property-Based Requirements (PBRs) are another mathematical formulation of system requirements based on a semilattice \citep{micouin2008toward, bernard2012requirements}. This formulation may be applied to other modeling languages such as AADL, Modelica, and VHDL-AMS, but \citet{micouin2008toward} focuses the presentation on SysML.
%<*mytagR1C2-1>
\SA{While PBRs focus on the integration and computation of physical properties to prepare for system V\&V, our work emphasizes the importance of well-formed requirement statements whose components are SysML model elements.}
%</mytagR1C2-1>
%<*mytagR4C3-4>
Like MBSR, PBR extends the standard SysML requirement element by creating a PBRequirement stereotype, and like MBSR, the PBRequirement stereotype provides four primary attributes representing 1) Condition, 2) Carrier, 3) Property, and 4) Domain \citep{micouin2008toward}. What is considered a well-formed PBR refers to its completion of these attributes to form a mathematical constraint, which in light of \citet{GtWR2023} is insufficient for mature requirements engineering practice. PBR does not distinguish between stakeholder needs and system requirements, whereas INCOSE-derived MBSR support\SA{s needs and need sets in addition to requirement expressions and requirement sets}. Composite requirements, hierarchical nesting of requirements, and automated identification of dependencies are features of PBR that MBSR supports using SysML (e.g. Fig. \ref{fig:RelationMap}).
\citet{SysML1.7} presents PBR as a non-normative extension with examples in Annex E, but this formulation differs from \citet{micouin2008toward} in that it focuses on ``quantitative specification of numerical parameters, relationships, equations and/or constraints.'' PBRs, as presented in the SysML v1.7 specification Annex E.8, may be adapted to the MBSR Profile to provide formal verification capabilities especially suited for numerical- and logic-focused requirements while benefitting from the INCOSE GtWR ontology and rules. \citet{lu2008requirement} presents another property-based formulation of requirements using a custom object-oriented modeling tool, and discusses similar benefits to MBSR, such as enhanced traceability, consistency, completeness, maintainability, and integration with artifact and report generation.
%</mytagR4C3-4>

Ontology-based requirements engineering is an active area of MBSE research with promising results that may contribute to the ASoT with reusable ontology-defined terms \citep{avdeenko2016ontology,yang2019ontology,lorch2024formal}. An ontology-based requirement may be composed of slots referring to Pattern Slots of the Requirement Statement, and the Attributes contributing to the complete Requirement Expression. Through establishing ontology relationships and axioms, model-based requirements V\&V may be assisted with tool automation ensuring (C10) completeness, (C11) consistency, (C15) correctness, (C13) ``unambiguity'', and (C14) ``traceability'' \citep{avdeenko2016ontology}. While the INCOSE-derived MBSR Profile may appear to implement an ontology, it lacks the formal semantics defined by ontology languages such as the Web Ontology Language (OWL) \citep{OWL2}. Therefore, the MBSR Profile is not currently capable of being exported to a standard ontology format, although a mapping from the meta-model to an ontology may be technically feasible. Integration with ontology tools may enhance the reasoning capability of a SysML model, but at the cost of tooling complexity, which MBSR attempts to avoid.

\subsection{Semantic Extensions to Model the Problem Space}
\label{sec:SemanticExtensions}

Alternatively, semantic extensions to SysML may be employed to model the problem space, known as True Model-Based Requirements (TMBR) \citep{salado2019constructing}. TMBR derives from the Wymorian theory of requirements (Sec. \ref{sec:MathBased}) and so ``attempts to model any type of requirement as a set of (or sets) of required input/output transformations'' \citep{salado2023model}. TMBR is not intended for modeling stakeholder needs \citep{salado2023model}, compared to INCOSE-derived MBSR which includes Needs and Need Sets in its meta-model \SA{(Fig. \ref{fig:Metamodel})}. The set theory foundations of TMBR prevent ``formal flaws in problem formulation, such as enforcing design solutions or leaving the requirement unbounded'' \citep{salado2023model}, corresponding to Characteristics (C15) Correct, (C12) Feasible and (C14) Able to be validated. Due to the reuse of SysML features to model the problem space, stakeholders may find TMBR difficult to understand compared to traditional requirement statements \citep{salado2024comparative}.
%<*mytagR4C3-5>
TMBR and MBSR are similar in that they use SysML as the primary modeling language and make use of SysML attributes with defined quantities and constraints. 
%<*mytagR1C3-1>
\SA{As a language extension, MBSR is compatible with other SysML profiles including SYSMOD \citep{weilkiens2020sysmod}.}
%</mytagR1C3-1>
%</mytagR4C3-5>

\section{Conclusion}
\label{sec:Conclusion}

\SA{While it is too early to determine the economic value of using INCOSE-derived MBSRs compared to traditional requirements engineering methods, initial results indicate the technique had a positive impact on a real project. We now present limitations of the present work, highlight paths forward for future work, and conclude with summarizing remarks.}

%%%%%%%%%%%%%%%%%%%%%%%%%%%%%%%%%%%%%%%%%%%%%%%%%%%%%%

\subsection{Limitations}
\label{sec:Limitations}

%<*mytagR1C4>
Limitations of the present work are primarily due to scoping and available SysML v1 tool capabilities. \SA{While our experiments have primarily made use of Cameo Systems Modeler, we posit that non-normative extensions to SysML v1 such as tables, matrices, and scripting capabilities are not restricted to certain tools, and that other tools will work in similar ways that we cannot fully describe here. Therefore, the MBSR technique is presented as a SysML extension, rather than a tool-grafted technique.}
%</mytagR1C4>
%<*mytagR5C29>
Although experimental design as done by \citet{salado2024comparative} to compare this MBSR approach with other tool-enabled requirements engineering methods was not conducted, qualitative validation of this MBSR approach was conducted in the context of two real-world system development projects at NASA Jet Propulsion Laboratory by the first author. %
%</mytagR5C29>
%<*mytagR3C17-3>
Producing experimental evidence of the MBSR contribution toward satisfying the 15 Characteristics of Well-formed Requirements and Sets of Requirements (Table \ref{tab:Characteristics}) is left as future work; readers are encouraged to refer to \citet{GtWR2023} and accompanying manuals available for free to INCOSE members.
%</mytagR3C17-3>
Contractual agreements currently limit the exposure of proprietary information such as the roughly 500 MBSRs written, so a minimal example of an asteroid sample collection spacecraft was used for demonstration. Although included in the MBSR Profile and linked to the applicable INCOSE GtWR Attribute sets, Needs and Need Sets were not used and evaluated in the projects. We provide the MBSR SysML profile in an open-source online repository \citep{herber2024repo} so other organizations and researchers may adapt it and run experiments to further the systems engineering community's understanding of the effectiveness of this approach.

\subsection{Future Work}
 \label{sec:FutureWork}

%<*mytagR1C2-2>
Future work will continue to explore the feasibility and effectiveness of INCOSE-derived MBSR. \SA{Future work remains to make full use of SysML modeling tool capabilities in conjunction with other tools such as MATLAB, Modelica, and Python to exploit the SysML-based traceability that the MBSR technique provides.}
%</mytagR1C2-2>
%<*mytagR2C4>
Some INCOSE GtWR Rules, such as those disallowing certain words and phrases, may be automatically checked using simple string matching, achieving similar capability to existing requirements quality management tools.
Such a capability may be encoded in automated validation suites using custom scripts similar to the ones we wrote for metrics and legends, although the potential performance impact is a concern addressable by future research.
Macros may be created that employ automated validation suite results to add and remove \Stereotype{Satisfy} and \Stereotype{Violate}
relationships for Rules amenable to automation.
%<*mytagR3C20>
\SA{Many other model validation rules may be added that further enhance the model-based requirements V\&V automation afforded by MBSR, such as checking that architecture decomposition levels match across MBSR-referenced model elements, or checking for conflicting or redundant requirements.}
%</mytagR3C20>
%</mytagR2C4>

%<*mytagR2C2>
Automatically-generated requirement statements, as proposed by \citet{herber2022model}, are provided as a read-only derived attribute, but future work is needed to make this `Derived Requirement Statement' attribute more readable while not interfering with model element naming conventions: such as by adding a stereotype to MBSR-linked elements that provides text attributes for defining the requirement statement fragment depending on its intended pattern slot. Names of the Pattern Slots and their element values could be used with Large Language Models or other advanced Natural Language Processing models to generate natural language sentences with proper syntax and grammar.
%</mytagR2C2>
The Condition pattern slot may be further decomposed following \citet{mavin2010big}, which categorizes conditions into event-driven, unwanted behavior, state-driven, and optional feature; or it may be decomposed according to a hierarchical ontology as shown in \citet{carson2015implementing} and the GtWR Appendix C.
The capacity of an ASoT with MBSRs to answer targeted mission programmatic questions is an area of future research that depends on the high maturity of both the system architecture model and requirements engineering organizational processes.

The revised semantics in SysML v2 present an opportunity to adapt the MBSR Profile to the new language and to research the change impact and potential benefits. SysML v2 uniformly applies the modeling language design pattern of \textit{definition} and \textit{usage} to requirements; it defines requirements necessarily as constraints with boolean satisfiability; stakeholder concerns are more explicitly modeled compared to SysML v1; and metadata elements are used instead of stereotypes. \SA{Furthermore, SysML v2 specifies a textual syntax in addition to the graphical syntax, potentially making model representations such as Fig. \ref{fig:XMI} more readable and amenable to automation.} SysML v2 also provides a standardized API for accessing the system architecture model, which may ease future research in integrating MBSR with NLP-based and formal requirements engineering tools. This shift toward more model-based requirements presents new opportunities to enrich the ASoT with metadata and automated model validation derived from the INCOSE GtWR.

\subsection{Summary}
\label{sec:Summary}

This paper presented an extension of prior MBSR research using the
latest INCOSE Guide to Writing Requirements and relate\SA{d} it \SA{to} an
emerging DRE paradigm. Over 300 requirements were written and revised
with project stakeholder feedback to test the effectiveness of this approach in improving quality and connectedness, contributing to a valuable ASoT.
The experience gained through this NASA JPL system development project was shared in brief \SA{in Sec.~\ref{sec:Discussion}}.
% while presenting a minimal example in the space flight domain.
The benefits and challenges listed in subsections \ref{sec:Benefits} and \ref{sec:Challenges}
\SA{were} gathered through experience, and while the benefits are likely transferable to other tools, the challenges
may be ameliorated by improved software performance and enhanced model-based workflows for DRE activities.

%<*mytagR1C2-3>
\SA{We found that development of SysML requirements in this direction has been fruitful for gaining support of MBSE in the context of requirements engineering.}
%</mytagR1C2-3>
By encoding the primary object classes (Fig.~\ref{fig:Metamodel}) from the INCOSE GtWR into the systems architecture modeling tool, rapid improvement in real-world requirements quality was achieved by a systems engineering student intern over a period of 6 months.
The INCOSE-derived \SA{MBSR profile} supports a DE approach
to requirements engineering and system development that may reduce costs and improve quality if
deployed at scale.
%<*mytagR1C3-2>
\SA{Our MBSR technique is independent of a methodology, as its intent is to extend the usefulness of SysML models that include requirements, from early-phase requirements development to architecture solution development to later-phase SE activities. The technique is therefore adaptable to new and existing SE methodologies.}
%</mytagR1C3-2>
SysML-based architecture modeling tools may facilitate the MBSR approach by supporting requirements management needs such as automatic timestamps, organization-defined collaborative workflows, extensible and rigorous requirement identifier definition, text values with embeddable model elements, and enhanced speed of the software.
The authors believe that \SA{INCOSE-derived MBSR has} the potential to leverage SysML strengths in the transition to DE while effectively satisfying stakeholder needs, but further \SA{development}, testing, feedback, and experimentation are needed to further validate this claim.

\section*{Acknowledgements}

The first author would like to thank Paulo J. Younse for his mentorship, and the Mars Returned Sample Handling team for their participation and support.

\section*{Conflict of Interest}
All authors declare that they have no conflicts of interest.

\section*{Supporting Information}

Access the open-source SysML profiles and example MBSR models on GitHub \citep{herber2024repo}.

%\printendnotes

\bibliographystyle{unsrtnat}
\bibliography{references}

%\clearpage
\vfill

\begin{biography}[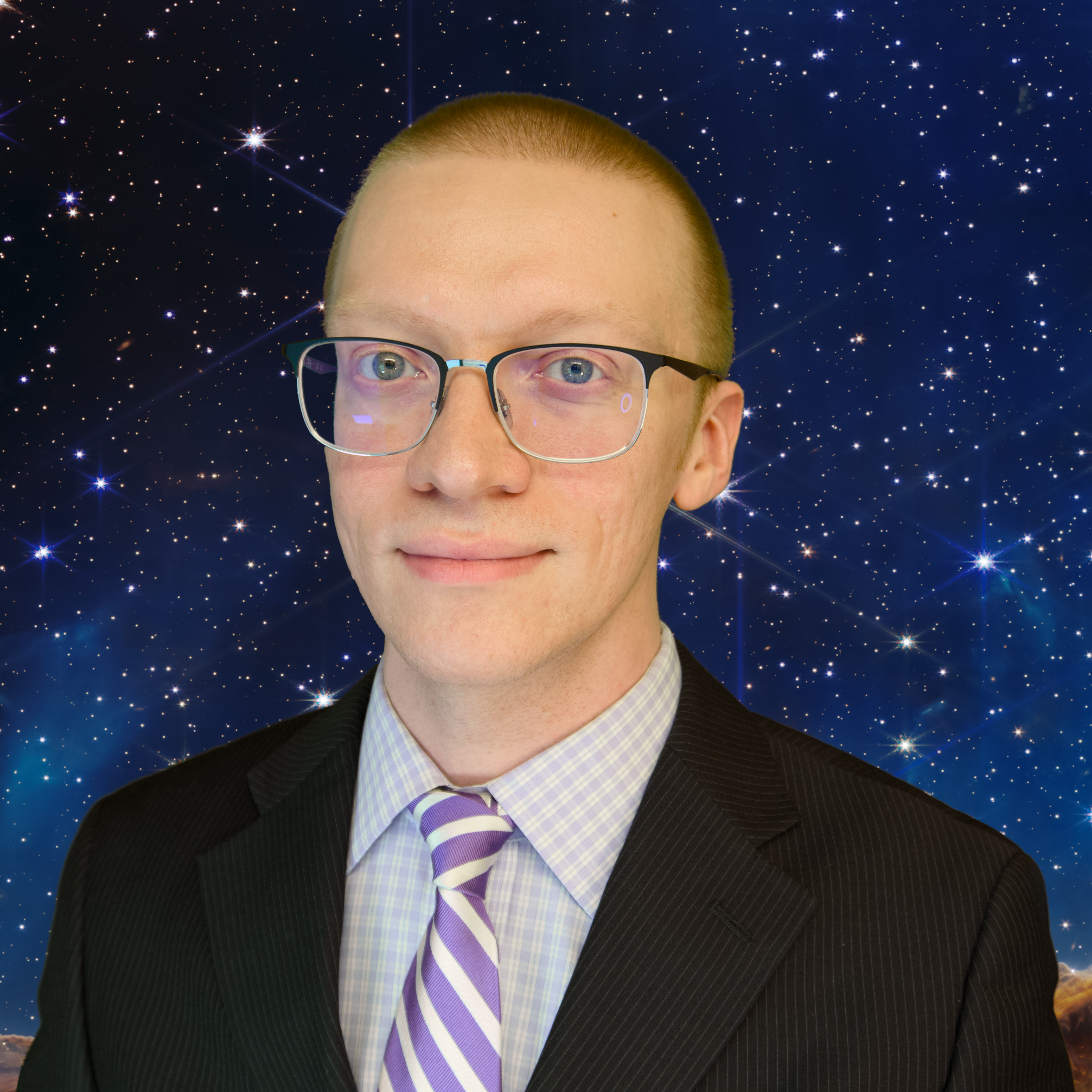]{James~S.~Wheaton}
is a Principal Systems Engineer at Terumo Blood and Cell Technologies in Denver, CO, USA. His research interests include digital engineering, model-based systems engineering, requirements engineering, ontology engineering, and provably-correct systems. James holds a B.S. in Mechanical Engineering from Purdue University (2011) and Ph.D. in Systems Engineering from Colorado State University. His professional background is in software engineering of ecommerce, big data, AI, and blockchain systems. James interned at NASA Jet Propulsion Laboratory from 2023-2024, where he developed system architectures and requirements for two Mars Sample Return projects. He holds certifications as Associate Systems Engineering Professional (ASEP) from INCOSE, and OMG-Certified Systems Modeling Language Professional (OCSMP) Model Builder Fundamental from the Object Management Group.
% \bigskip
% \bigskip
\end{biography}
\begin{biography}[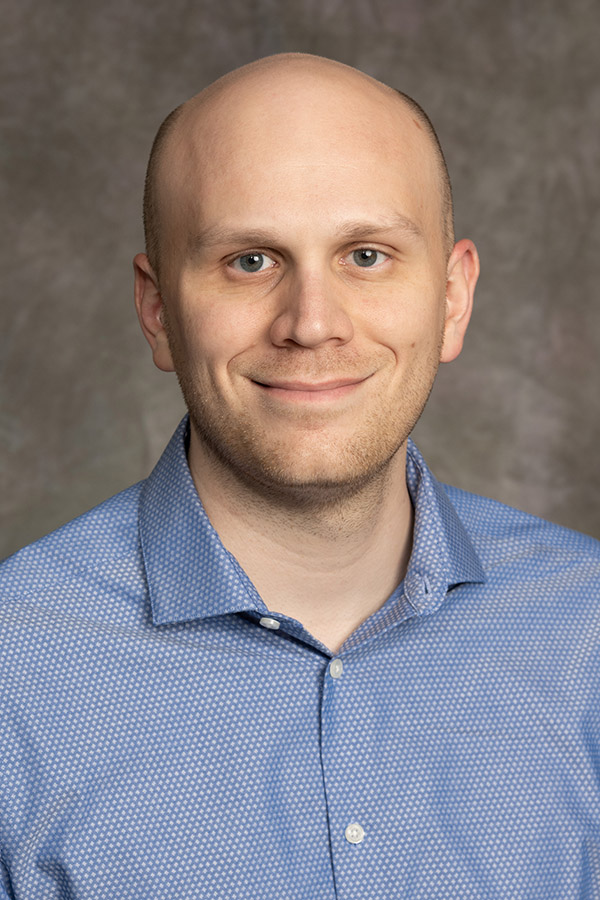]{Daniel~R.~Herber}
is an Associate Professor in the Department of Systems Engineering at Colorado State University in Fort Collins, CO, USA. His research interests and projects have been in design optimization, model-based systems engineering, system architecture, digital engineering, dynamics and control, and combined physical and control system design (control co-design), frequently collaborating with academia, industry, and government laboratories. His work has involved several application domains, including energy, aerospace, defense, and software systems. He teaches courses in model-based systems engineering, system architecture, controls, and optimization. He is a member of INCOSE, ASME, and AIAA.
He studied at the University of Illinois at Urbana-Champaign, earning his B.S. (2011) in General Engineering and his M.S. (2014) and Ph.D. (2017) in Systems and Entrepreneurial Engineering. He held a postdoctoral position (2018-2019) with the NSF ERC for Power Optimization for Electro-Thermal Systems (POETS).
% \bigskip
% \bigskip
\end{biography}

\clearpage
\appendix
\section{Appendix}

%%%%
\begin{figure}[h]
\includegraphics[width=\columnwidth]{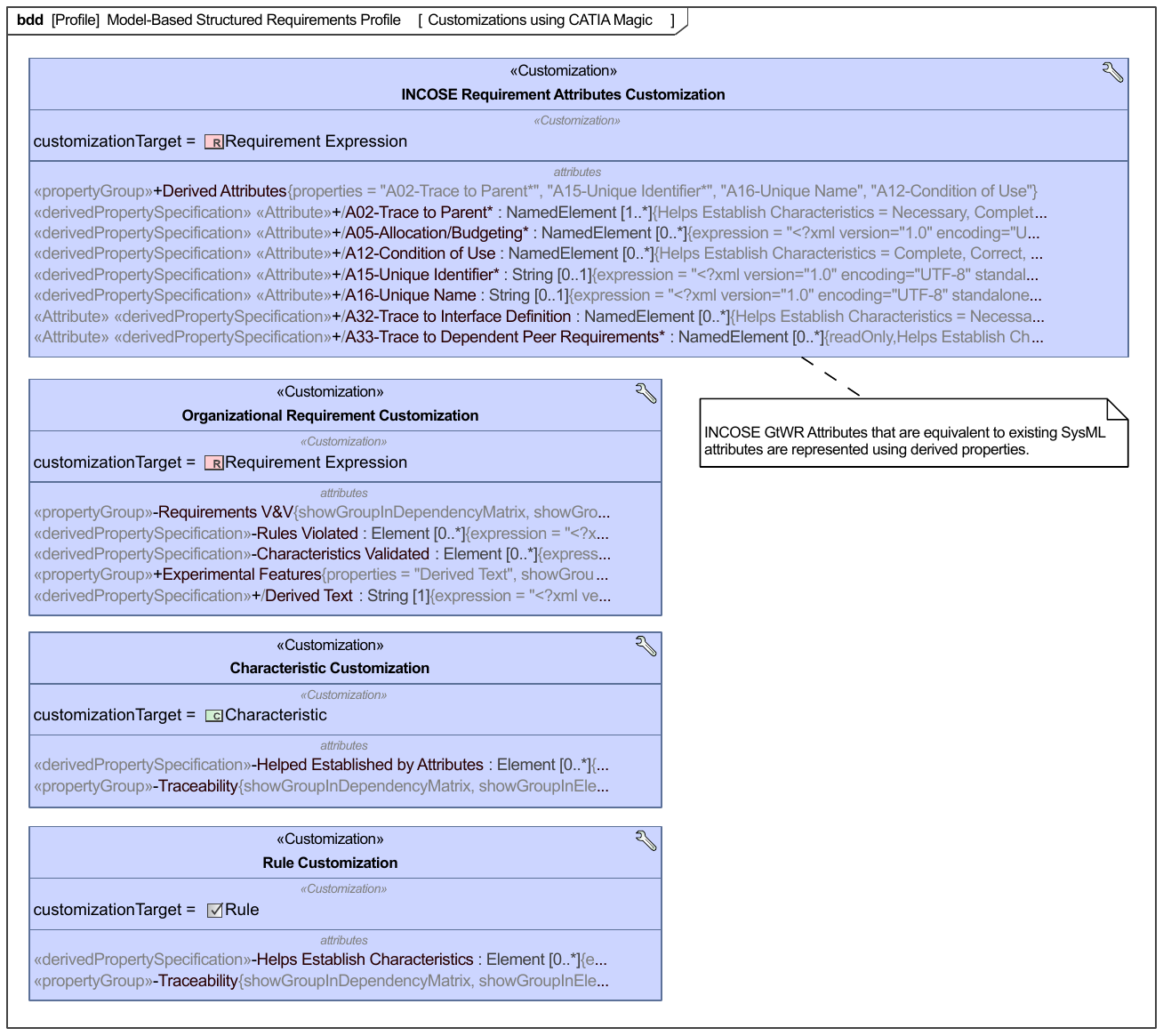}
\caption{\SB{Customizations of the INCOSE-derived MBSR attributes (Fig.~\ref{fig:GtWR-Attributes}) using Cameo Systems Modeler facilities.}}
\label{fig:Customizations}
\end{figure}
%%%%

\end{document}